\documentclass[%
aps
 amsmath,amssymb,
preprint,%
]{revtex4-1}

\usepackage{subfigure}
\usepackage{graphicx}% Include figure files
\usepackage{dcolumn}% Align table columns on decimal point
\usepackage{bm}% bold math
\usepackage{appendix}
\begin{document}

%\preprint{APS/123-QED}

\title{Electron dynamics in counter-propagating laser waves} % Force line breaks with \\

\author{Yanzeng Zhang}
 %\email{yaz148@ucsd.edu}
\affiliation{%
 Mechanical and Aerospace Engineering Department, University of California San Diego, La Jolla, CA 92093, USA
}%\\
\author{Sergei Krasheninnikov}%
 %\email{skrash@mae.ucsd.edu}
\affiliation{%
 Mechanical and Aerospace Engineering Department, University of California San Diego, La Jolla, CA 92093, USA
}%


\begin{abstract}

The electron dynamics in counter-propagating laser waves is investigated by employing a novel approach, where the new Hamiltonian is time-independent when the perturbative laser wave is absent. The physical picture of stochastic electron dynamics is clearly revealed and the threshold values of the amplitude of the perturbative laser field for triggering stochastic electron acceleration are derived for different laser polarization directions and initial electron momentum. It demonstrates that the dephasing rate (new Hamiltonian) between the electron and the dominant laser can be randomly reduced if the amplitude of the perturbative laser is above the threshold such that the electron could be accelerated by the dominant laser well beyond the ponderomotive energy scaling. The impact of a superluminal phase velocity is examined, which slightly changes the stochastic region in Hamiltonian space if the superluminal phase velocity is under a threshold value but significantly decreases the maximum electron kinetic energy. All the analytic considerations are confirmed by the numerical simulations.

\end{abstract}

\maketitle
\section{Introduction}
The generation of energetic electrons through laser-plasma interaction, which relies on energy conversion from the laser radiation to electrons, has been the subject of many theoretical and experimental studies due to their potential applications. Many mechanisms of electron acceleration have been proposed, including laser wake-field acceleration \cite{rosenbluth1972excitation,tajima1979laser,wagner1997electron,santala2001observation,geddes2004high} and direct laser acceleration with the assistance of different configurations of quasi-static electromagnetic fields \cite{smith1975stochastic,*menyuk1987stochastic,pukhov1999particle,gahn1999multi,tanimoto2003direct,arefiev2012parametric,paradkar2011numerical,*krasheninnikov2014stochastic,tsakiris2000laser}, etc. 

One kind of direct laser acceleration is to consider electrons in multiple laser pulses \cite{forslund1985two,bauer1995relativistic,shvets1998superradiant,mendonca1982stochasticity,*mendonca1983threshold,sheng2002stochastic,*sheng2004efficient,bourdier2005stochastic,bochkarev2018stochastic}, where the mechanism of electron acceleration can be attributed to an onset of stochasticity when the amplitudes of the perturbing (weaker) laser waves exceed some threshold values \cite{mendonca1983threshold,sheng2002stochastic,bourdier2005stochastic,bochkarev2018stochastic} and the maximum electron energy can be well beyond the ponderomotive scaling. It was shown that the most efficient stochastic heating is to consider two counter-propagating laser waves \cite{sheng2004efficient,bourdier2005stochastic}. Such configuration of two laser beams can be due to the reflection of the dominant incident laser beam from the target surface \cite{sentoku2002high} or Raman backscattering. 

However, due to the multidimensional spatio-temporal characteristics of the laser waves and strong nonlinearity of the dynamics of relativistic electrons in these waves, the analytic investigations of stochastic electron acceleration in the colliding laser waves in earlier studies are rather limited and the approaches used are quite complicated \cite{zaslavskii1968stochastic,rechester1979stochastic,LDLandaubook,mendonca1982stochasticity,*mendonca1983threshold,rax1992compton,bourdier2005stochastic}, while the numerical simulations, which can shed some light on the criterion for stochastic electron motion in multiple laser waves \cite{sheng2002stochastic,*sheng2004efficient,bochkarev2018stochastic,bourdier2005stochastic}, are only valid within the simulated parameter range. Therefore, more complete theoretical analysis is needed to have a better understanding of the electron dynamics in the counter-propagating laser waves.

Recently, it was shown that, by employing the integrals of motion for electrons in laser and quasi-static electromagnetic fields, the electron dynamics can be described by the 3/2 dimensional (3/2D) Hamiltonian approach \cite{angus2009energy,zhang2018electron1,*zhang2018electron2}, which has greatly simplified the analysis of electron dynamics \cite{zhang2018stochastic,zhang2018electronlongitudinalE} and the boundary of electron energy due to the stochastic motion was obtained by finding the Chirikov-like mapping \cite{chirikov1979universal}. Such method has been extended to the case of electrons in the colliding laser waves \cite{zhang2019stochastic} by employing proper canonical variables and effective time, such that the new Hamiltonian becomes time independent when the perturbative laser wave is absent, where the electron dynamics for luminal planar laser waves, which are linearly polarized in the same direction, and transverse canonical momentum being zero was exhaustively examined. It demonstrated that the electron energy gained from a relativistic laser wave via the stochastic acceleration due to the presence of a perturbative counter-propagating laser wave can greatly exceed the ponderomotive energy scaling, where the essential role of the perturbation is to change the dephasing rate (new Hamiltonian) between the electron and dominant laser.  

This work is devoted to extending investigation of electron dynamics in the colliding laser waves by considering different laser polarization directions, initial electron canonical momentum and an impact of superluminal phase velocity. Following \cite{mendonca1982stochasticity,rax1992compton,bourdier2005stochastic,sheng2002stochastic,zhang2019stochastic}, we will consider the case where one of the counter-propagating laser waves is much stronger than the other one, which thus can been taken as a perturbation.

The remainder of this paper is organized as follows. The new Hamiltonian equations will be introduced in section \ref{sec_Hamil_traj} and the unperturbed electron trajectories are examined. Section \ref{sec_stochastic_threshold} will investigate the conditions for stochasticity for different laser polarization directions and initial electron momentum. An impact of the superluminal phase velocity will be discussed in section \ref{sec_superl} and numerical simulations by directly integrating electron equations of motion will be presented in section \ref{sec_simulation}. The main results will be summarized and discussed in section \ref{sec_conclusions}.

\section{New Hamiltonian equations and unperturbed electron trajectories}

In this section, we will summarize the main idea of how to find the new Hamiltonian \cite{zhang2019stochastic} and then examine the unperturbed electron trajectories in this new framework. To this purpose, in the following, the dimensionless variables will be used, where $\mathbf{r}$ and $t$ are normalized, respectively, by the dominant laser wavenumber $k$ and $kc$ ($c$ is the speed of light in vacuum). The normalized parameter of the laser wave, described by a vector potential $\mathbf{A}$, is $e\mathbf{A}/mc^2$, where $-e$ and $m$ are the electron charge and mass.

We assume that the dominant laser wave propagates along $z$ direction and is described by the vector potential of $\mathbf{A}(v_pt-z)$ ($v_p\geq 1$ is the normalized phase velocity), whereas the perturbative laser wave is described by $\mathbf{A}_1(v_pt+z)$, both of which are arbitrarily polarized in $x$ and $y$ directions. Then the electron dynamics can be described by the Hamiltonian:
\begin{equation}
\mathcal{H}\equiv\gamma=\left[1+(\mathbf{P}+\mathbf{A}+\mathbf{A_1})\right]^{1/2},\label{eqHamil-general}
\end{equation}
where $\gamma$ is the relativistic factor and $\mathbf{P}=\gamma \mathbf{v}-\mathbf{A}-\mathbf{A_1}$ is the canonical momentum. It is easy to show that for such laser field, the canonical momentum in $x,y$ directions (denoting as $\bar{P_x}$ and $\bar{P}_y$) are conserved so that the Hamiltonian in equation~(\ref{eqHamil-general}) is effectively two dimensional. From equation~(\ref{eqHamil-general}), we can introduce new variables of $\eta=v_pt-z$, $\chi=\gamma+v_pP_z$ and time $\tau= v_pt+z$, such that the electron dynamics can be described by a new Hamiltonian $H(\chi,\eta,\tau)\equiv \gamma-v_pP_z$. One can show that $\eta=v_pt-z$ and $\chi=\gamma+v_pP_z$ are canonical variables provided that
\begin{equation}
\eta d\chi-Hd\tau=2(P_zdz-\mathcal{H} dt),\label{eqcanonical_trans}
\end{equation}
is hold \cite{landau1978course}. The new Hamiltonian can be found as 
\begin{equation}
H(\chi,\eta,\tau)=\frac{2v_p}{v_p^2-1}\sqrt{\chi^2+(v_p^2-1)P_\perp^2}-\frac{v_p^2+1}{v_p^2-1}\chi,\label{eqHamiltonian}
\end{equation}
where $P_\perp^2=1+\sum_{i=x,y}\left[\bar{P}_i+A_i(\eta)+A_{1i}(\tau)\right]^2$, while the Hamiltonian equations read
\begin{equation}
\frac{d\chi}{d\tau}=\frac{\partial H}{\partial \eta},~\textup{and}~\frac{d\eta}{d\tau}=-\frac{\partial H}{\partial \chi}.\label{eqnewHalmileq}
\end{equation}
This new Hamiltonian equations, which can also be obtained from the electron equations of motion, will greatly simplify our analysis in comparison with the multidimensional Hamiltonian 
\cite{mendonca1983threshold,rax1992compton} based on equation~(\ref{eqHamil-general}).

For simplicity, we first consider the luminal case $v_p=1$, while the impact of superluminal phase velocity, which mimics the impact of plasma, will be qualitatively discussed in section \ref{sec_superl} and then numerically investigated in section \ref{sec_simulation}. The linearly polarized planar laser waves will be used in the following analysis, i.e., $\mathbf{A}=a sin(\eta)\mathbf{e}_x$ and $\mathbf{A}_1=a_1sin(k_1\tau)\mathbf{e}_x$ or $\mathbf{A}_1=a_1sin(k_1\tau)\mathbf{e}_y$ depending on the relative polarization directions of the counter-propagating waves, where $a_1\ll a$ and $k_1$ is the ratio of the perturbative laser frequency (or wavenumber) to that of the dominant one. Then, the Hamiltonian in equation~(\ref{eqHamiltonian}) degenerates to
\begin{equation}
H=\frac{1+\left[asin(\eta)+\delta_x a_1sin(k_1\tau)+\bar{P}_x\right]^2+\left[\delta_y a_1sin(k_1\tau)+\bar{P}_y\right]^2}{\chi},\label{eqHamiltonian_luminal}
\end{equation}
where $\delta_{x,y}=0~\textup{or}~ 1$ are switches to controlling the perturbative laser polarization direction. Keep in mind that we are interested in the gain of maximum electron kinetic energy, $\gamma_{max}$, which can be expressed in the terms of $H$, for $v_p=1$, as follows:
\begin{equation}
\gamma_{max}\equiv \frac{\chi+H}{2}\approx\frac{1}{2}\left(\frac{E_p}{H}+H\right), \label{eqgamma_max}
\end{equation}
 where $E_{p}=1+(a+|\bar{P}_x|)^2+\bar{P}_y^2$. Note that the ponderomotive scaling for pre-accelerated electron in the dominant laser wave only is $E_p/H_0$, where $H_0$ is the conserved dephasing rate (which corresponds to the initial Hamiltonian in the present problem). Therefore, $\gamma_{max}$ can significantly exceed the ponderomotive scaling either for $H_{min}<H_0$ (which corresponds to the electron moving along the dominant laser propagation direction and energy gain ratio is $H_0/H_{min}$) or for $H_{max}>E_p/H_0$ (where the electron moves along the perturbative laser propagation direction and energy gain ratio is $H_{max}H_0/E_p$). The latter case requires $k_1\gg a\gg 1$ \cite{zhang2019stochastic}, which is not considered in this paper.

For the unperturbed problem ($a_1=0$), the new Hamiltonian is conserved and from equations~(\ref{eqnewHalmileq}, \ref{eqHamiltonian_luminal}) we find the following implicit dependence $\eta(\tau)$ (we note that $\eta$ increases with $\tau$ provided $d\eta/d\tau>0$):
\begin{equation}
\tau=\frac{2\bar{P}^2+a^2}{4H^2}\left[2\eta-\frac{a^2sin(2\eta)}{2\bar{P}^2+a^2}-\frac{8a\bar{P}_xcos(\eta)}{2\bar{P}^2+a^2}\right]+const.,\label{eq-tau-eta}
\end{equation}
where $\bar{P}^2=1+\bar{P}_x^2+\bar{P}_y^2$; and $\chi$ depending on time $\tau$:
\begin{equation}
\chi=\frac{1+\left[asin(\eta)+\bar{P}_x\right]^2+\bar{P}_y^2}{H}.\label{eqChi_tau}
\end{equation}
From equation~(\ref{eq-tau-eta}) one can find the frequency of unperturbed oscillation of electron canonical coordinate $\chi$:  
\begin{equation}
\omega=\frac{2\pi}{T}=\frac{2H^2}{2\bar{P}^2+a^2},\label{eqOmega}
\end{equation}
where $T=\tau(\eta=2\pi)-\tau(\eta=0)$ is the period of electron oscillation. Therefore, the presence of $\bar{P}_{x,y}$ will decrease (increase) the frequency (period) of electron oscillation via $\bar{P}^2$ and alter the electron trajectories as shown in equations~(\ref{eq-tau-eta}, \ref{eqChi_tau}). 

\begin{figure}[bt]
\centering
\begin{minipage}{1\linewidth}
\includegraphics[width=0.6\textwidth]{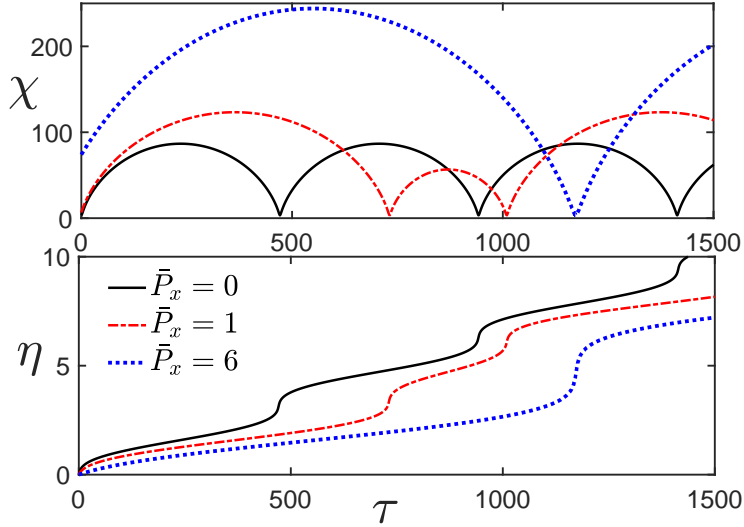}
\caption{Schematic view of electron trajectories for $a=5$, $k_1=1$, $\bar{P}_y=0$, and different $\bar{P}_x$. For $\bar{P}_x=0~\textup{or}~ 1$, $H=0.3$ and for $\bar{P}_x=6$, $H=0.5$.}
\label{fig-electron-trajectories}
\end{minipage}
\end{figure}

From equation~(\ref{eqHamiltonian_luminal}) one can find that, for relativistic case $a> 1$ (which we will consider in the following), unperturbed (or weakly perturbed) electron trajectories have characteristics of zig-zag time dependence of canonical coordinate $\chi$ (e.g., see the upper panel of figure~\ref{fig-electron-trajectories}). This feature of electron trajectories enables a long tail of the distribution of the amplitude of harmonics, making stochastic electron motion possible. Also, from equation~(\ref{eqHamiltonian_luminal}) it follows that the strongest impact, ``kicks'', on both $H$ and canonical variables by the perturbative laser occurs at a very short time near the local minimum of $\chi$ (e.g., see figure~\ref{fig-H_diff-chi}), where the phase between electron and backward laser wave is locally minimized and $\eta$ undergoes jump \cite{zhang2019stochastic}. The positions of minima of $\chi$ depend both on $\bar{P}_x$ and $a$: when $|\bar{P}_x|<a$, $\chi$ is minimized at $\eta_1=(2n+1)\pi+\phi$ and $\eta_2=2n\pi-\phi$, where $\phi=sin^{-1}(\bar{P}_x/a)$ and $n$ is an integer; whereas, for $\bar{P}_x>a$ ($\bar{P}_x<-a$), the minima of $\chi$ are obtained only at $\eta_3=-\pi/2+2n\pi$ ($\eta_4=\pi/2+2n\pi$).

\label{sec_Hamil_traj}

\section{Threshold for stochastic electron motion}

\label{sec_stochastic_threshold}

The unperturbed electron motion could resonate with the perturbative laser wave when $m\omega=k_1$ (where $m$ is the harmonics of unperturbed electron motion) as seen from equation~(\ref{eqHamiltonian_luminal}). When $\omega\ll k_1$, overlapping of the separatrices of neighbouring resonant islands, $\bar{K}=(\delta \omega+\delta \omega')/2\Delta \omega>1$, where
$\delta \omega$ and $\delta \omega '$ are their widths and $\Delta\omega$ is the distance between them, is possible and stochastic heating occurs \cite{roal1990nonlinear}. The condition for an onset of stochasticity for the case $\omega/k_1\ll 1$ can also be found by using equivalent, but more convenient Chirikov-like mapping deduced from electron equations of motion, which will be mainly used in the following.

\begin{figure}[bt]
\centering
\begin{minipage}{1\linewidth}
\includegraphics[width=0.6\textwidth]{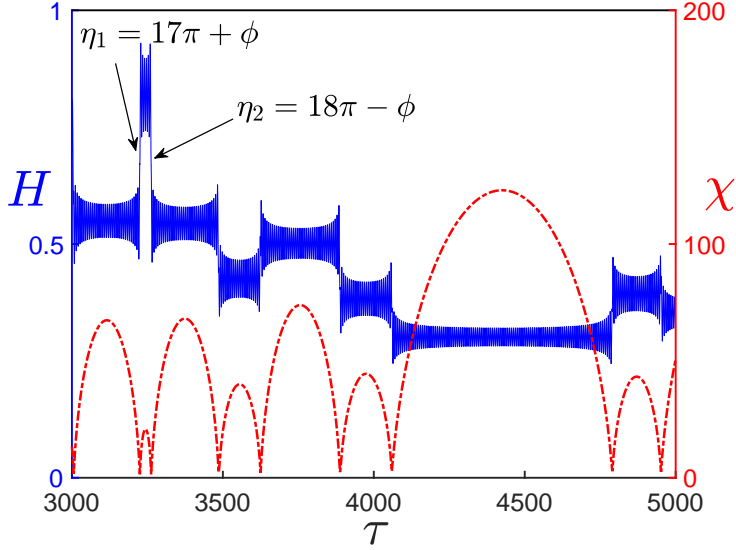}
\caption{Schematic view of diffusion of Hamiltonian (solid blue) and corresponding $\chi$ (dash-dot red) for $a=5$, $a_1=0.2$, $k_1=1$, $\bar{P}_y=0$, and $\bar{P}_x=1$.}
\label{fig-H_diff-chi}
\end{minipage}
\end{figure}

As discussed in the last section, the kicks due to the perturbative laser of $H$ takes place at a short time near the local minimum of $\chi$ (e.g., see figure\ref{fig-H_diff-chi}). Except these short periods of time $\tau$ where $\eta\approx \eta_{1,2}$ for $|\bar{P}_x|<a$ and $\eta\approx \eta_{3}~(\eta_4)$ for $\bar{P}_x>a~(\bar{P}_x<-a)$, the electron ``sees" only fast phase change of the backward laser wave due to large $\chi=\gamma+P_z$ and, therefore, undergoes adiabatic oscillation. Therefore, the Chirikov-like mapping can be formed by using the Hamiltonian $H_n$ and time $\tau_n$, when the electron passes through the nonadiabatic region. Such mapping corresponds to the Poincar\'e section of electron crossing effectively ``fixed" canonical momentum ($\eta$) plane. Let's assume that the change of the Hamiltonian due to each nonadiabatic interaction of electron with the perturbative wave is smaller than the Hamiltonian itself, i.e., $\Delta H=\left|H_{n+1}-H_n\right|\ll H_n$, then the unperturbed electron trajectory $H_n(\eta,\chi)$ can be used to estimate the variation of Hamiltonian due to each kick \cite{zaslavskii1987stochastic}:
\begin{equation}
\Delta H_n\equiv H_{n+1}-H_n=\int_{\eta\approx \eta_i}\frac{\partial H}{\partial \tau}d\tau,~i=1,2~\textup{or}~3~\textup{or}~4.\label{eqDelta_H}
\end{equation}

\subsection{Counter-propagating waves with same polarization direction}

We first consider the case where the perturbative laser is polarized along the dominant one, i.e., $\mathbf{A}_1=a_1 sin(k_1\tau)\mathbf{e}_x$ such that $\delta_x(\delta_y)=1(0)$ in equation~(\ref{eqHamiltonian_luminal}). Under the condition of $a_1\ll a$, we could keep the leading term of $\partial H/\partial \tau=2a_1k_1\left[asin(\eta)+\bar{P}_x\right]cos(k_1\tau)/\chi$. The fact that the main contribution to Hamiltonian variation is from $\eta\approx \eta_i,~i=1,2,3,4$ enables us to do the expansion of the integrand in equation~(\ref{eqDelta_H}) with respect to $\eta-\eta_i$. 
For the case of $|\bar{P}_x|<a$, we have 
\begin{eqnarray}
\frac{\Delta H_n}{H_n}&=&\pm \frac{2a_1\beta^{1/2}}{(1+\bar{P}_y^2)^{1/2}}sin(k_1\tau_n) \label{eqDelta_H_2}\int_{-\infty}^{\infty}\tilde{\eta}sin\left(\beta\tilde{\eta}+\frac{1}{3}\tilde{\eta}^3\right)d\tilde{\eta},
\end{eqnarray}
where $\tilde{\eta}=(\eta-n\pi)/\alpha$, $\alpha=(H_n^2/k_1a^2cos^2\phi)^{1/3}\sim(\omega/k_1)^{1/3}\ll 1$, $\beta=(k_1/H_n^2a)^{2/3}(1+\bar{P}_y^2)/cos^{2/3}\phi$, and the ``+" (``-") sign denotes the variation of $H$ at $\eta_1$ ($\eta_2$). It should be noted that the fast oscillation for $\tilde{\eta}~\tilde{>}1$ justifies the extension of the integration limits to infinity and the nonadiabatic interaction of electron motion with backward laser occurs at $|\eta-n\pi|<\alpha\ll 1$ ($|\tilde{\eta}|\tilde{<}1$).

The integral in equation~(\ref{eqDelta_H_2}) could be expressed with the derivative of Airy function, $Ai'(\beta)$, so we have
\begin{equation}
\Delta H_n=\pm \frac{4\pi a_1\beta^{1/2}Ai'(\beta)}{(1+\bar{P}_y^2)^{1/2}}H_nsin(\psi_n),\label{eqDelta_H_3}
\end{equation}
where $\psi\equiv k_1\tau_n$. Taking into account the properties of the Airy function, it follows that the requirement of $\Delta H<H_n$ is always satisfied for $a_1\tilde{<}1$. 

If $\bar{P}_x=0$, the time interval between $\eta_1$ and $\eta_2$ is half of the electron oscillating period, i.e., $\tau(\eta_1)-\tau(\eta_2)=T/2$, such that the phase interval between two consecutive kicks is determined by the Hamiltonian:
\begin{equation}
\Delta \psi_n\equiv\psi_{n+1}-\psi_n=k_1T/2=\frac{\pi k_1\left[2(1+\bar{P}_y^2)+a^2\right]}{2H_{n+1}^2}.\label{eqDelta_xi_final}
\end{equation}
As a result, rearranging equations~(\ref{eqDelta_H_3}, \ref{eqDelta_xi_final}) could form symplectic mapping conserving phase volume and allow us to find the condition for an onset of stochasticity
\begin{equation}
K_x\equiv\left|\frac{d\Delta \psi_n}{dH_{n+1}}\frac{d\Delta H_n}{d\psi_n}\right|=\frac{4\pi^2 aa_1\left[2(1+\bar{P}_y^2)+a^2\right]\beta^2|Ai'(\beta)|}{(1+\bar{P}_y^2)^{2}}\tilde{>}1.\label{eqK_general}
\end{equation}
It follows that the presence of $\bar{P}_y$ only provides a factor less than unity and, therefore, will increase the threshold of $a_1$ ($K_x\approx 1$) for triggering stochastic motion, whereas it doesn't change the basic features of stochastic electron dynamics. This is because $\bar{P}_y$ simply increases the effective electron mass. Then, following \cite{zhang2019stochastic} by using the property of $f(\beta)\equiv 4\pi^2\beta^2|Ai'(\beta)|$, we find that the stochastic acceleration is only possible for $a_1>a_{sx}$, where
\begin{equation}
a_{sx}\approx\frac{0.11(1+\bar{P}_y^2)^{2}}{a\left[2(1+\bar{P}_y^2)+a^2\right]},\label{eqThreshold-a2}
\end{equation}
and the stochastic acceleration occurs in the vicinity of $H\approx H_{sx}$ ($\beta\approx 1.68$ for $f(\beta)$ reaching its maximum), where 
\begin{equation}
H_{sx}\approx 0.68(1+\bar{P}_y^2)^{3/4} \left(\frac{k_1}{a}\right)^{1/2}.\label{eqThreshold-H}
\end{equation}
The expressions for lower and upper boundaries of the stochastic region \cite{zhang2019stochastic} remain unchanged with the presence of $\bar{P}_y$:
\begin{equation}
H_{min}^x\approx\frac{H_{sx}}{\sqrt{1.6+0.69ln\left(a_1/a_{xs}\right)}},~\textup{and}~H_{max}^x\approx 1.5\left(\frac{a_1}{a_{sx}}\right)^{3/8}H_{sx}, \label{eqH_boundary_Ax}
\end{equation}
where $H_{min}^x$ has a weak logarithmic dependence on the ratio $a_1/a_{sx}>1$. However, given that $a_{sx}$ and $H_{sx}$ increase with $\bar{P}_y$, we find that the lower (upper) boundary of stochastic region increases (decreases) with the presence of $\bar{P}_y$. Therefore, the ratio of the maximum electron kinetic energy over the ponderomotive scaling decreases as seen from equation~(\ref{eqgamma_max}). 

However, an impact of $\bar{P}_x$ on the stochastic condition is more complex providing that it not only changes the effective electron mass but also increases the energy exchange between electron and laser through the work done by the laser electric field in $x$-direction. Considering that the time interval from $\eta_1$ to $\eta_2$ is different from that from $\eta_2$ to next $\eta_1$ for $0<|\bar{P}_x|<a$ (e.g., see figure~\ref{fig-electron-trajectories}), the method of Chirikov-like mapping to find the stochastic condition is not convenient, but we could resort to the method of island overlapping. As shown in Appendix \ref{AppendixVmn}, we could find the stochastic condition as
\begin{equation}
\bar{K}^2=a_1\frac{16m^2}{(2\bar{P}^2+a^2)}\sum_{h=0,\pm 1}\left[ha+2\delta_h^0\bar{P}_x\right]C_{m-h}\left [\frac{ma^2}{2(2\bar{P}^2+a^2)},\frac{4ma\bar{P}_x}{2\bar{P}^2+a^2}\right]\tilde{>}1,\label{eqSquare_of_K}
\end{equation}
where $m\equiv k_1/\omega$ is the resonant harmonics, $\delta_i^j$ is the Kronecker symbols, and $C_N(\alpha,\beta)$ is the generalized Bessel function \cite{rax1992compton,nikishov1964ai}. Notice that similar results were obtained in \cite{rax1992compton, bochkarev2018stochastic} by using multidimensional Hamiltonian methods. For $\bar{P}_x=0$, one can show that $K_x\approx \bar{K}^2$.

\begin{figure}[ht]
\centering
\begin{minipage}{0.6\linewidth}
\includegraphics[width=1\textwidth]{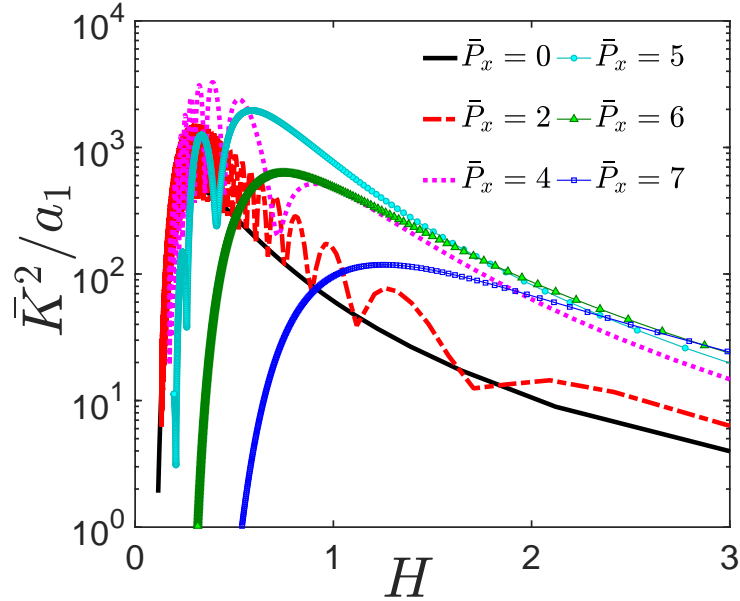}
\caption{Schematic view of $g\left[m(H)\right]=\bar{K}^2/a_1$ versus $H$ for different $\bar{P}_x$, where $a=5$, $k_1=1$, and $\bar{P}_y=0$. }
\label{fig-f-m-px}
\end{minipage}
\end{figure}

Then, we can define $\bar{K}^2\equiv a_1g(m)$, where $g(m)$ shows how $\bar{K}^2$ varies with different $m$ and thus $H$ (e.g., see figure~\ref{fig-f-m-px}). As a result, the threshold of stochastic instability requires $a_1>(max\{g(m)\})^{-1}$. As we can see from figure~\ref{fig-f-m-px}, for $|\bar{P}_x|\tilde{<}a$, the maximum value of $g(m)$ increases with the presence of $\bar{P}_x$ and thus the threshold value obtained in equation~(\ref{eqThreshold-a2}) for $\bar{P}_x=0$ decreases. Whereas for $|\bar{P}_x|\gg a$, the peak of $g(m)$ decreases with increasing $|\bar{P}_x|$, which is eventually smaller than that for $\bar{P}_x=0$ meaning that the effect of increasing the effective electron mass becomes dominant. Notice that there could be multiple stochastic peaks for $|\bar{P}_x|\tilde{<} a$ and from equation~(\ref{eqSquare_of_K}) the result is symmetric with respect to $\bar{P}_x=0$. On the other hand, we notice that both the lower and upper stochastic boundaries in $H$ ($g\sim 1/a_1$) shifts toward larger $H$. Therefore, the ratio of the maximum electron kinetic energy gained from stochasticity over the ponderomotive scaling, $H_0/H_{min}^x$ for $k_1\sim 1$, will decrease with the presence of $\bar{P}_x$ (although $H_{max}^x$ increases with $|\bar{P}_x|$, the energy gain, $H_{max}H_0/E_p$ for $k_1\gg 1$, will also decreases given that $E_P$ grows faster than $H_{max}^x$). 

\subsection{Counter-propagating waves with orthogonal polarization directions}

Then we consider the case where the colliding laser waves have orthogonal polarization directions, i.e., $\mathbf{A}_1=a_1sin(k_1\tau)\mathbf{e}_y$ such that $\delta_y(\delta_x)=1(0)$ in equation~(\ref{eqHamiltonian_luminal}). For this case, we have $\partial H/\partial \tau=2a_1k_1\left[a_1sin(k_1\tau)+\bar{P}_y\right]cos(k_1\tau)/\chi$, where we kept the second order term of $a_1^2sin(k_1\tau)cos(k_1\tau)$ for general $\bar{P}_y$. In the following, we take $\bar{P}_x=0$. Similar to the former case, we can do expansion of $\tau$ with respect to $\eta-\eta_i$ when estimating the Hamiltonian variation in the nonadiabatic region. As a result, we have
\begin{equation}
\Delta H_n=\frac{2\pi a_1^2\beta Ai(2^{2/3}\beta)}{(1+\bar{P}_y^2)}H_nsin(2\psi_n)+\frac{4\pi a_1\bar{P}_y\beta Ai(\beta)}{(1+\bar{P}_y^2)}H_nsin(\psi_n),\label{eqDelta_H_Ay}
\end{equation}
whereas the time interval between two consecutive kicks is given in equation~(\ref{eqDelta_xi_final}).

For $\bar{P}_y=0$, we see that the phase in equation~(\ref{eqDelta_H_Ay}) corresponding to Chirikov-like mapping is twice that in equation~(\ref{eqDelta_H_3}) for $\mathbf{A}_1$ polarized along with $\mathbf{A}$. As a result, the stochastic condition can be found from equations~(\ref{eqDelta_H_3}, \ref{eqDelta_H_Ay}) as
\begin{equation}
K_y=4\pi^2 aa_1^2(2+a^2)\beta^{5/2}|Ai(2^{2/3}\beta)|\tilde{>}1.\label{eqK_general-Ay}
\end{equation}
Introducing the function $f_2(\beta)=4\pi^2\beta^{5/2}|Ai(2^{2/3}\beta)|$, we find that $f_2(\beta)$ first increases with $\beta$ for $\beta<\beta_{sy}\approx 1.10$; reaches its maximum, $f_{max}\approx 2.55$, at $\beta_{sy}$; and then falls exponentially at $\beta>\beta_{sy}$ (e.g., see Ref. \onlinecite{gradshteyn2014table}) as
\begin{equation}
f_2(\beta)\approx 2^{5/6}\pi^{3/2}\beta^{9/4}exp\left[-(4/3)\beta^{3/2}\right].\label{eqf_largebeta}
\end{equation}
One can show that for $\beta<\beta_{sy}$, $f_2$ can also be approximated by the expression (\ref{eqf_largebeta}) except a factor of order of unity. As a result, from equation~(\ref{eqK_general-Ay}) we find that stochastic acceleration is only possible for $a_1>a_{sy}$, where
\begin{equation}
a_{sy}= \left[\frac{f_{max}^{-1}}{a(2+a^2)}\right]^{1/2}\approx\frac{0.63}{\left[a(2+a^2)\right]^{1/2}},\label{eqThreshold-a2-Ay}
\end{equation}
and the stochastic acceleration occurs in the vicinity of $H\approx H_{sy}$ ($\beta\approx \beta_{sy}$), where 
\begin{equation}
H_{sy}\approx 0.93 \left(\frac{k_1}{a}\right)^{1/2}.\label{eqThreshold-H-Ay}
\end{equation}
It follows that the threshold in equation~(\ref{eqThreshold-a2-Ay}) is larger than that in equation~(\ref{eqThreshold-a2}) for the case of parallel polarized laser waves. 

For $a_1\gg a_{sy}$, stochastic acceleration becomes possible within the range of $H$: $H_{min}^y<H<H_{max}^y$, where $H_{max}^y$ and $H_{min}^y$ could be found by using equation~(\ref{eqf_largebeta}) of the function $f_2(\beta)$. Notice that the inequalities $a\gg a_1\gg a_{sy}$ could be satisfied for $a\gg 1$, under which we obtain:
\begin{equation}
H_{min}^y\approx\frac{H_{sy}}{\sqrt{1.56+1.30ln\left(a_1/a_{sy}\right)}}, \label{eqH_minimal}
\end{equation}
and
\begin{equation}
H_{max}^y\approx 1.35\left(\frac{a_1}{a_{sy}}\right)^{2/3}H_{sy}, \label{eqH_maxim}
\end{equation}
where we have taken the numeric factor into account when using equation~(\ref{eqf_largebeta}) for $\beta<\beta_{sy}$. Considering that $a_{sy}$ is much lager than $a_{sx}$, $H_{min}^y$ ($H_{max}^y$) is relatively lager (smaller) than $H_{min}^x$ ($H_{max}^x$) for the same $a$ and $a_1$ ($\bar{P}_{x,y}=0$). Therefore, the maximum electron kinetic energy in equation~(\ref{eqgamma_max}) for two lasers being orthogonally polarized is smaller than that for two lasers being parallel polarized when $\bar{P}_{x,y}=0$.

On the other hand, if $|\bar{P}_y|\gg a_1$, the variation of Hamiltonian in equation~(\ref{eqDelta_H_Ay}) is mainly determined by the second term and the stochastic condition reads 
\begin{equation}
\bar{K}_y=\frac{4\pi^2 a_1a\bar{P}_y\left[2(1+\bar{P}_y^2)+a^2\right]\beta^{5/2}|Ai(\beta)|}{(1+\bar{P}_y^2)^{5/2}}\tilde{>}1.\label{eqK_general-Ay-Py}
\end{equation}
As a result, we could find the stochastic condition by using $f_3(\beta)=4\pi^2\beta^{5/2}|Ai(\beta)|$ as:
\begin{equation}
a_1>\bar{a}_{sy}= \frac{0.12(1+\bar{P}_y^2)^{5/2}}{a\bar{P}_y\left[2(1+\bar{P}_y^2)+a^2\right]},\label{eqThreshold-a2-Ay-Py}
\end{equation}
and the most unstable Hamiltonian:
\begin{equation}
\bar{H}_{sy}\approx 0.66(1+\bar{P}_y^2)^{3/4} \left(\frac{k_1}{a}\right)^{1/2}.\label{eqThreshold-H-Py}
\end{equation}
It follows that the threshold in equation~(\ref{eqThreshold-a2-Ay-Py}) is smaller than that in equation~(\ref{eqThreshold-a2-Ay}) for $\bar{P}_y=0$ if $|\bar{P}_y|\tilde{<} a^{3/8}$ and even comparable with $a_{sx}$ when $|\bar{P}_y|\tilde{>} 1$. However, considering that the lower boundary of stochastic region has a weak dependence on $a_1/\bar{a}_{sy}$: 
\begin{equation}
\bar{H}_{min}^y\approx\frac{\bar{H}_{sy}}{\sqrt{1.68+0.65ln\left(a_1/\bar{a}_{sy}\right)}}, \label{eqH_minimal_py}
\end{equation}
the increase of $\bar{H}_{sy}$ with $\bar{P}_y$ will make $\bar{H}_{min}^y$ above $H_{min}^y$ for $\bar{P}_y=0$ in equation~(\ref{eqH_minimal}). As a result, the ratio of the maximum electron kinetic energy against the ponderomotive scaling, $H_0/H_{min}$, decreases. The upper boundary of stochasticity is obtained as
\begin{equation}
\bar{H}_{max}^y\approx 1.32\left(\frac{a_1}{\bar{a}_{sy}}\right)^{1/3}\bar{H}_{sy}, \label{eqH_maxim_py}
\end{equation}
which is above $H_{max}^y$ providing that both $\bar{P}_y$ and $a_1/\bar{a}_{sy}$ have larger values.

\section{Impact of superluminal phase velocity}

In this section, we examine the impact of superluminal phase velocity, $v_p >1$, on the stochastic electron dynamics assuming that $\bar{P}_{x,y}=0$. For this purpose, we should again consider the unperturbed electron trajectories with conserved $H$, which, from equation~(\ref{eqHamiltonian}), reads
\begin{equation}
\chi= \frac{2v_p\sqrt{H^2+(v_p^2-1)(1+a^2sin^2\eta)}}{v_p^2-1}-\frac{(v_p^2+1)H}{v_p^2-1}. \label{eqTrajectory-vp}
\end{equation}
Therefore, $\chi$ reaches its maximum and minimum at, respectively, $\eta=\pi/2+n\pi$ and $\eta=n\pi$ as
\begin{eqnarray}
\chi_{max}&=& \frac{2v_p\sqrt{H^2+(v_p^2-1)(1+a^2)}}{v_p^2-1}-\frac{(v_p^2+1)H}{v_p^2-1},\nonumber \\
\chi_{min}&=&\frac{2v_p\sqrt{H^2+v_p^2-1}}{v_p^2-1}-\frac{(v_p^2+1)H}{v_p^2-1}. \label{eqChi-extremes}
\end{eqnarray}
It follows that both $\chi_{max}$ and $\chi_{min}$ decrease with increasing $H$ and so is $\gamma_{max}=(\chi_{max}+H)/2\approx \chi_{max}/2$ (we consider $k_1\sim 1$ such that $\gamma_{max}$ is dominated by the first term). Therefore, the maximum electron kinetic energy is obtained at the lower boundary ($H_{min}$) of the stochastic region. Noticing that $\partial H/\partial \tau$ is maximized at $\chi_{min}$, we know that the strongest impact of the perturbative laser on electron motion for superluminal case also occurs at $\chi_{min}$ corresponding to $\eta\approx n\pi$ for $\bar{P}_x=0$. On the other hand, the period of electron oscillation is given by
\begin{equation}
T/2\pi=\frac{v_p^2+1}{v_p^2-1}-\frac{4v_p HK(b^2)}{\pi(v_p^2-1)\sqrt{H^2+(v_p^2-1)(1+a^2)}}, \label{eqPeriod-vp}
\end{equation}
where $K(b^2)\equiv\int_0^{\pi/2}\frac{d\eta}{\sqrt{1-b^2cos^2\eta}}$ is the complete elliptic integral of the first kind and $b^2=(v_p^2-1) a^2/[H^2+(v_p^2-1)(1+a^2)]$. 

Then when $H\tilde{>}\sqrt{(v_p^2-1)(1+a^2)}$ such that $K(b^2\ll 1)\approx (1+b^2/4)\pi/2$, the electron oscillation period $T$ in equation~(\ref{eqPeriod-vp}) is approximate to that for the luminal case, so are the extrema of $\chi$ in equation~(\ref{eqChi-extremes}). As a result, the electron trajectories and thus the variation of Hamiltonian $\Delta H$ in equation~(\ref{eqDelta_H}) remains almost unchanged compared with those of luminal case. Then the stochastic region with $\sqrt{(v_p^2-1)(1+a^2)}\tilde{<}H$ is not affected by the superluminal phase velocity. It follows that if the lower boundary of the stochastic region $H_{min}$ for the luminal case satisfies such condition, i.e., $\sqrt{(v_p^2-1)(1+a^2)}\tilde{<}H_{min}$, the impact of $v_p>1$ on both $H$ and electron kinetic energy is negligible.

On the other hand, when $H_{min}\ll \sqrt{(v_p^2-1)(1+a^2)}$, from equation~(\ref{eqPeriod-vp}) we see that $T$ has an approximately linear dependence on $H$ for $H\ll \sqrt{(v_p^2-1)(1+a^2)}$, so that $|d\Delta \psi_n/dH_{n+1}|\approx 12v_pk_1/(v_p^2-1)\sqrt{(v_p^2-1)(1+a^2)}$ is a constant where we use $K(b^2\rightarrow 1)=ln(4/\sqrt{1-b^2})\approx 3$ (notice that the dominant term of $T$ is the first one on the right hand side of equation~(\ref{eqPeriod-vp}) such that its value remains almost unchanged with $H$). The variation of $H$ in the nonadiabatic region for small $H$ can be estimated as $\Delta H\sim aa_1(v_p-1)^{1/6}$ from equations~(\ref{eqHamiltonian}, \ref{eqTrajectory-vp}). Then the stochastic parameter in equation~(\ref{eqK_general}) is a constant value as $K_s\sim a_1/(v_p-1)^{4/3}$ for $|H|\ll \sqrt{(v_p^2-1)(1+a^2)}$, where a factor of order of unity has been omitted. It follows that a threshold value of $v_p$, $v_{ps}-1 \sim a_1^{3/4}$, exists such that for $v_p<v_{ps}$, the region of $|H|\ll \sqrt{(v_p^2-1)(1+a^2)}$ is stochastic and the lower boundary of the stochastic region in luminal case can extend to negative $H\gg -\sqrt{(v_p^2-1)(1+a^2)}$ (for further negative $H$ the stochasticity is impossible since $\chi$ increases smoothly from $\chi_{min}$ such that the zig-zag temporal dependence of $\chi$ is eliminated); whereas for $v_p>v_{ps}$, the stochasticity in small $H$ region is terminated. The maximum electron kinetic energy for the latter case is rather limited, while for the former case ($v_p<v_{ps}$), taking into account that $\chi_{max}$ and thus $\gamma_{max}$ weakly depends on $H$ for $|H|\ll \sqrt{(v_p^2-1)(1+a^2)}$, it can be estimated as 
\begin{equation}
\gamma_{max}\approx  \chi_{max}(H=0)/2 =v_p\sqrt{(1+a^2)/(v_p^2-1)}.\label{eqgamma_max_super}
\end{equation}
It shows that the maximum electron kinetic energy for superluminal case is much smaller than that for the luminal case, $\gamma_{max}\approx E_p/2H_{min}$ as shown in equation~(\ref{eqgamma_max}), even though the stochastic regions in $H$ space are almost the same. Moreover, for the superluminal case, the amplitude of oscillation of electron in the adiabatic region, $\delta H\sim 2a_1\sqrt{v_p^2-1}$, is approximately a large constant, whereas for luminal case, it decreases with decreasing $H$. These conclusions are confirmed by the numerical simulations presented in the next section.

\label{sec_superl}

\section{Numerical simulations}

\label{sec_simulation}

To verify the results of our analytical considerations, we integrate equations~(\ref{eqHamiltonian}-\ref{eqHamiltonian_luminal}) numerically and present the results in the Poincar\'e maps of ($H$, $\psi$) or ($\gamma$, $\psi$), when $\eta=2n\pi+ \pi/2$, where $\chi$ and thus $\gamma=(\chi+H)/2$ reaches their maximum in one unperturbed electron period (here we use $\bar{P}_{x,y}\geq 0$). Notice that even though for $\bar{P}_x=0$ the unperturbed period is $\Delta \eta=\pi$, we use $\Delta \eta=2\pi$ instead in all the mappings considering the general case with $\bar{P}_x$).

\begin{figure}
\centering
\subfigure{
\label{fig-H-a-5-a1x-02-py-0}
\begin{minipage}[bht]{0.45\textwidth}
\includegraphics[width=1\textwidth]{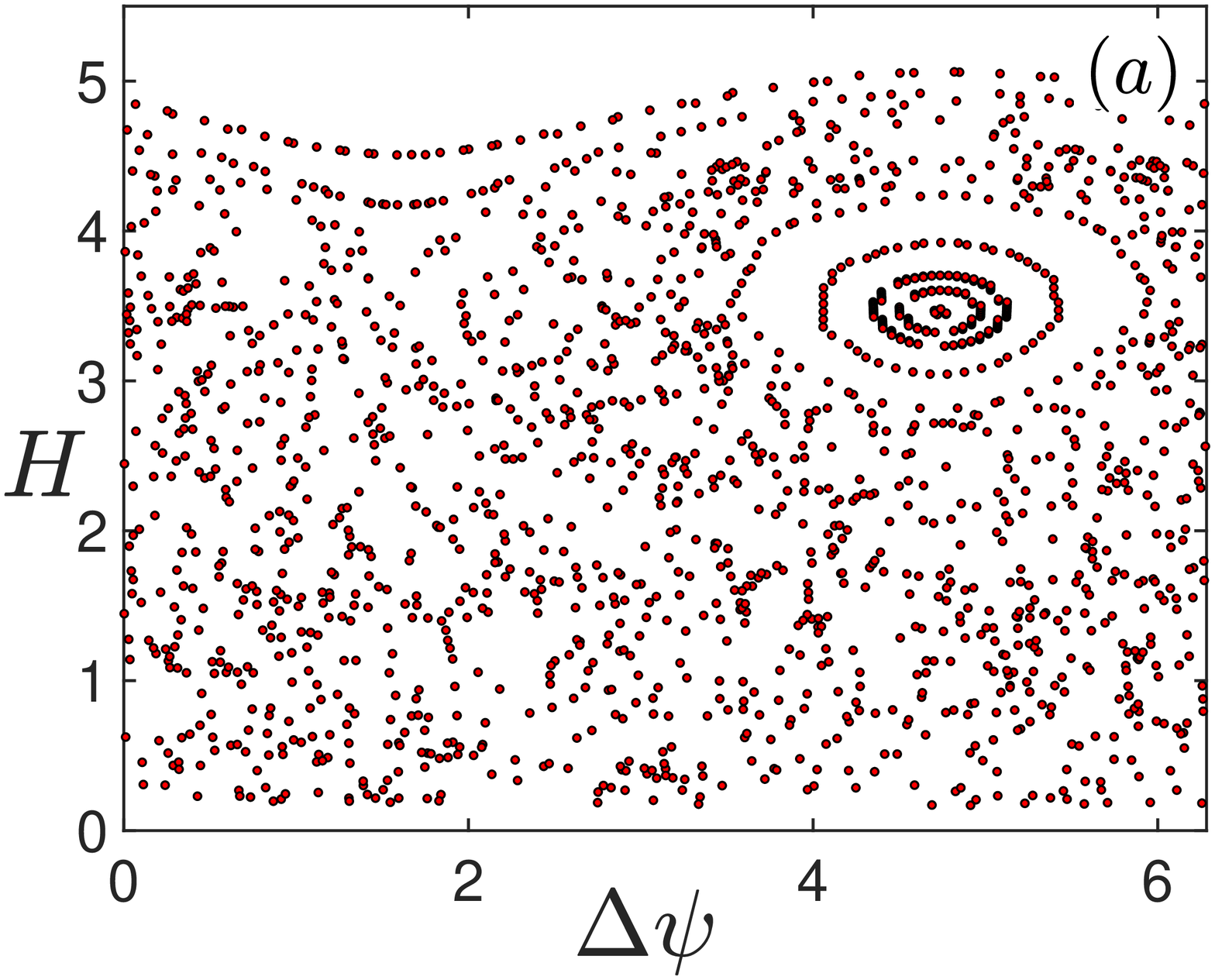}
\end{minipage}}
\subfigure{
\label{fig-H-a-5-a1x-02-py-2}
\begin{minipage}[bht]{0.45\textwidth}
\includegraphics[width=1\textwidth]{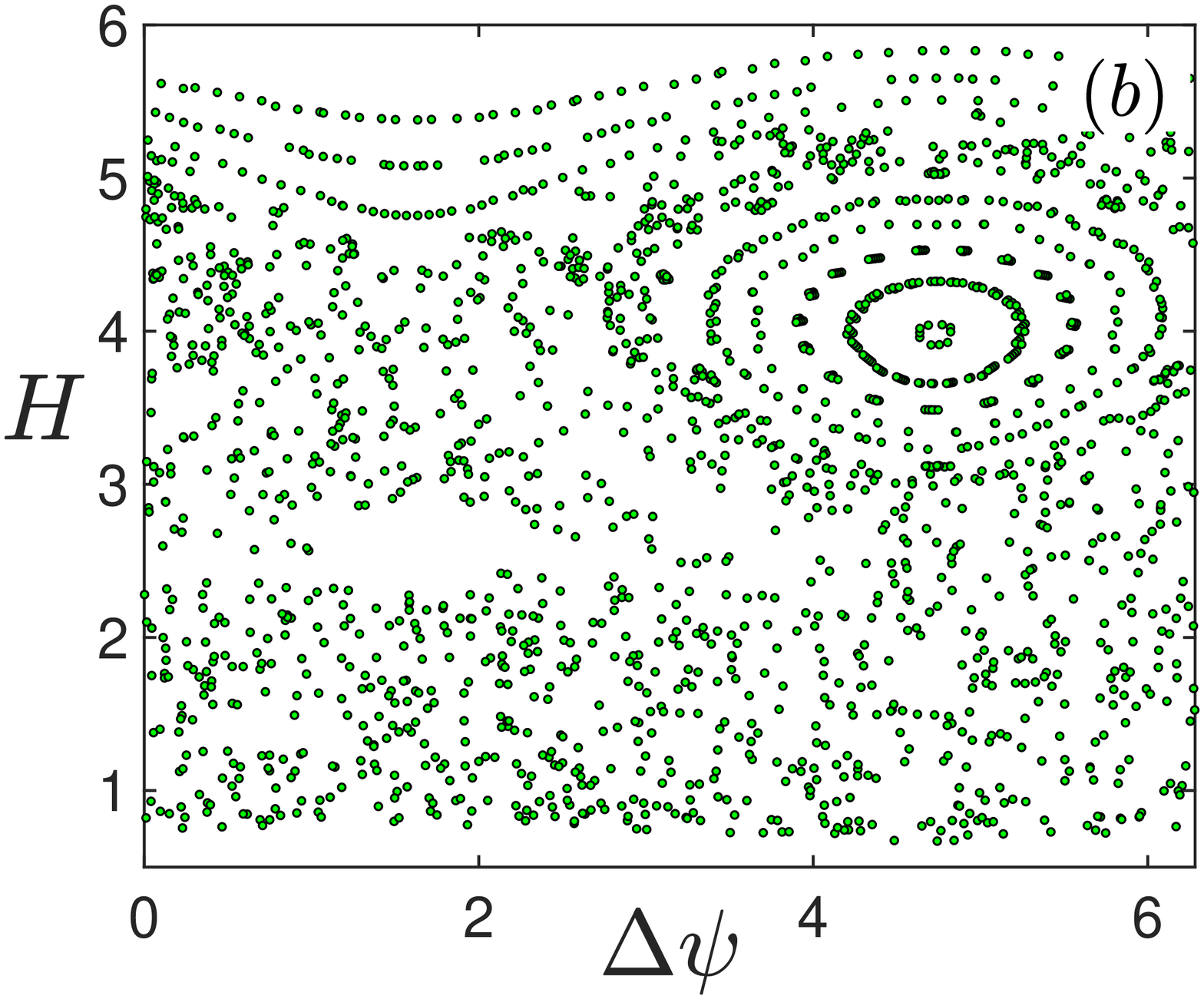}
\end{minipage}}
\subfigure{
\label{fig-H-a-5-a1x-0005-px-0}
\begin{minipage}[bht]{0.45\textwidth}
\includegraphics[width=1\textwidth]{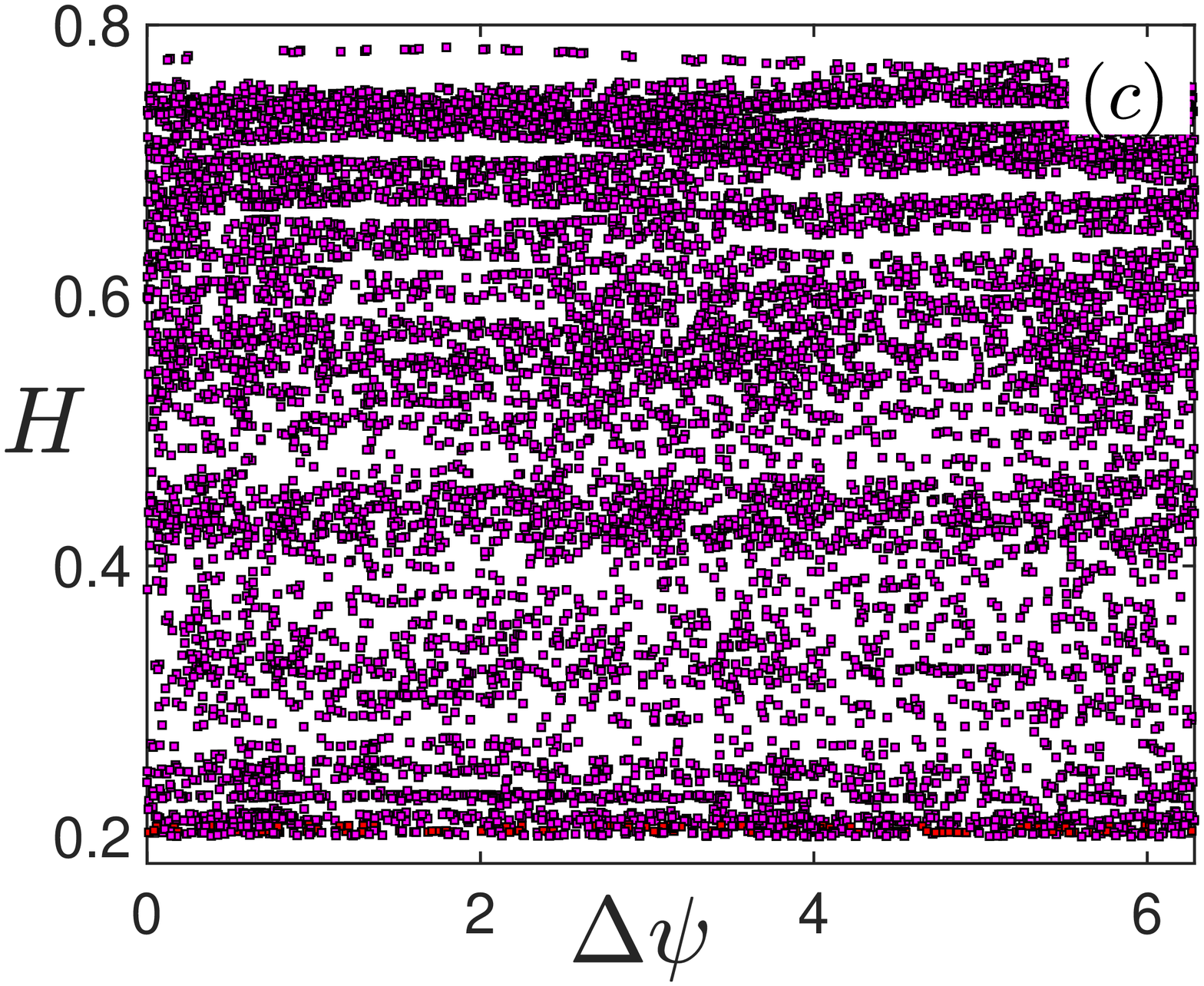}
\end{minipage}}
\subfigure{
\label{fig-H-a-5-a1x-0005-px-5}
\begin{minipage}[bht]{0.45\textwidth}
\includegraphics[width=1\textwidth]{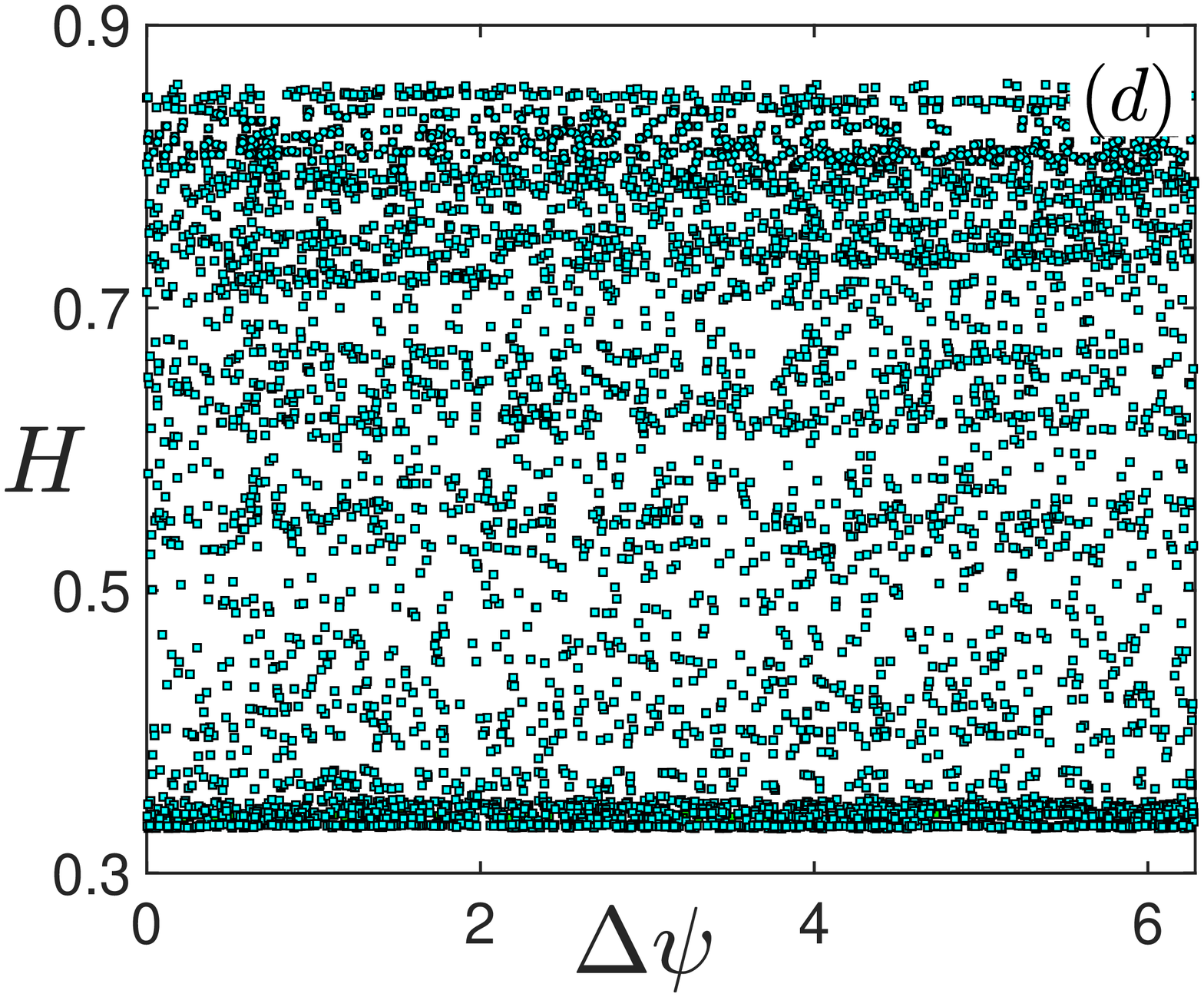}
\end{minipage}}
\caption{Poincar\'e mappings of ($H$, $\psi$) at $\eta=2n\pi+\pi/2$ for $a=5$, $\mathbf{A}_1=a_1sin(\tau)\mathbf{e}_x~(k_1=1)$ and different $a_1$, $\bar{P}_{x,y}$, where $\Delta \psi\equiv \psi-[\psi/2\pi]\times 2\pi$. (a) $a_1=0.2$, $\bar{P}_x=0$ and $\bar{P}_y=0$; (b) $a_1=0.2$, $\bar{P}_x=0$ and $\bar{P}_y=2$; (c) $a_1=0.005$, $\bar{P}_x=0$ and $\bar{P}_y=0$; (d) $a_1=0.005$, $\bar{P}_x=5$ and $\bar{P}_y=0$.  }
\label{figChirikov-Ax}
\end{figure}

Shown in figure~\ref{figChirikov-Ax} are the results for counter-propagating lasers with same polarization directions, where the parameters are $a=5$, $k_1=1$, and different $\bar{P}_{x,y}$ and $a_1$. As one can see, a stochastic ``sea" is bounded by the KAM invariant \cite{roal1990nonlinear} at $H_{min}^x$ and $H_{max}^x$, which fully agree with equation~(\ref{eqH_boundary_Ax}) for $\bar{P}_x=0$. As seen from figure~\ref{fig-H-a-5-a1x-02-py-0} and \ref{fig-H-a-5-a1x-02-py-2}, the presence of $\bar{P}_y$ increases the lower boundary of the stochastic region and thus decreases the energy gain ratio. Comparing figure~\ref{fig-H-a-5-a1x-02-py-0} with figure~\ref{fig-H-a-5-a1x-0005-px-0} where $a_1$ is different, we confirm that $H_{min}^x$ has a weak dependence on $a_1$ being above the threshold value $a_{sx}$. From figure~\ref{fig-H-a-5-a1x-0005-px-0} and \ref{fig-H-a-5-a1x-0005-px-5}, we see that both the lower and upper boundaries of the stochastic region become lager (while the corresponding maximum electron kinetic energy gain ratio becomes smaller) with the presence of $|\bar{P}_x|>0$ , which agrees with the prediction from figure~\ref{fig-f-m-px}.

\begin{figure}
\centering
\subfigure{
\label{fig-H-a-5-a1y-02-py-0}
\begin{minipage}[bht]{0.45\textwidth}
\includegraphics[width=1\textwidth]{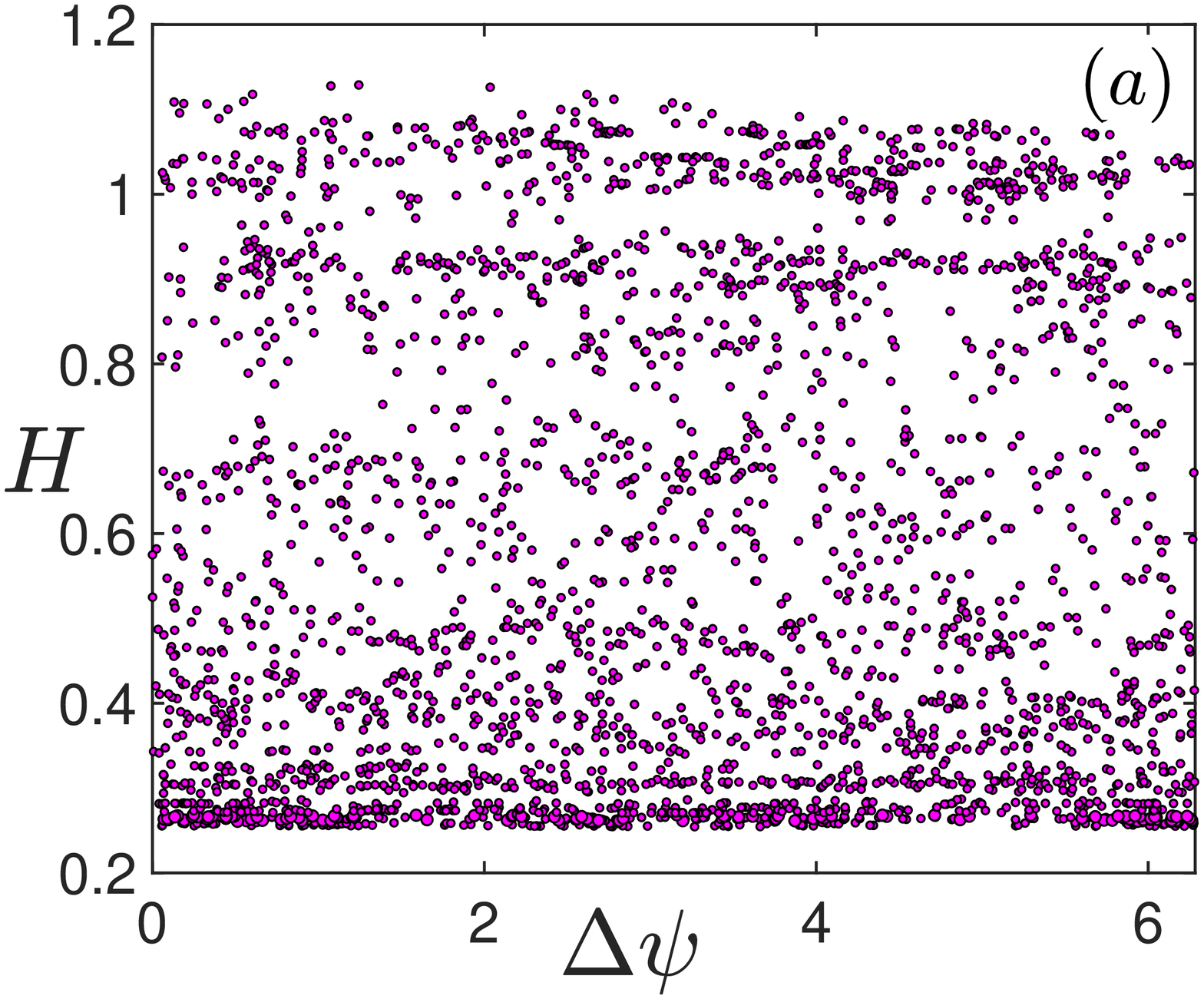}
\end{minipage}}
\subfigure{
\label{fig-H-a-5-a1y-02-py-2}
\begin{minipage}[bht]{0.45\textwidth}
\includegraphics[width=1\textwidth]{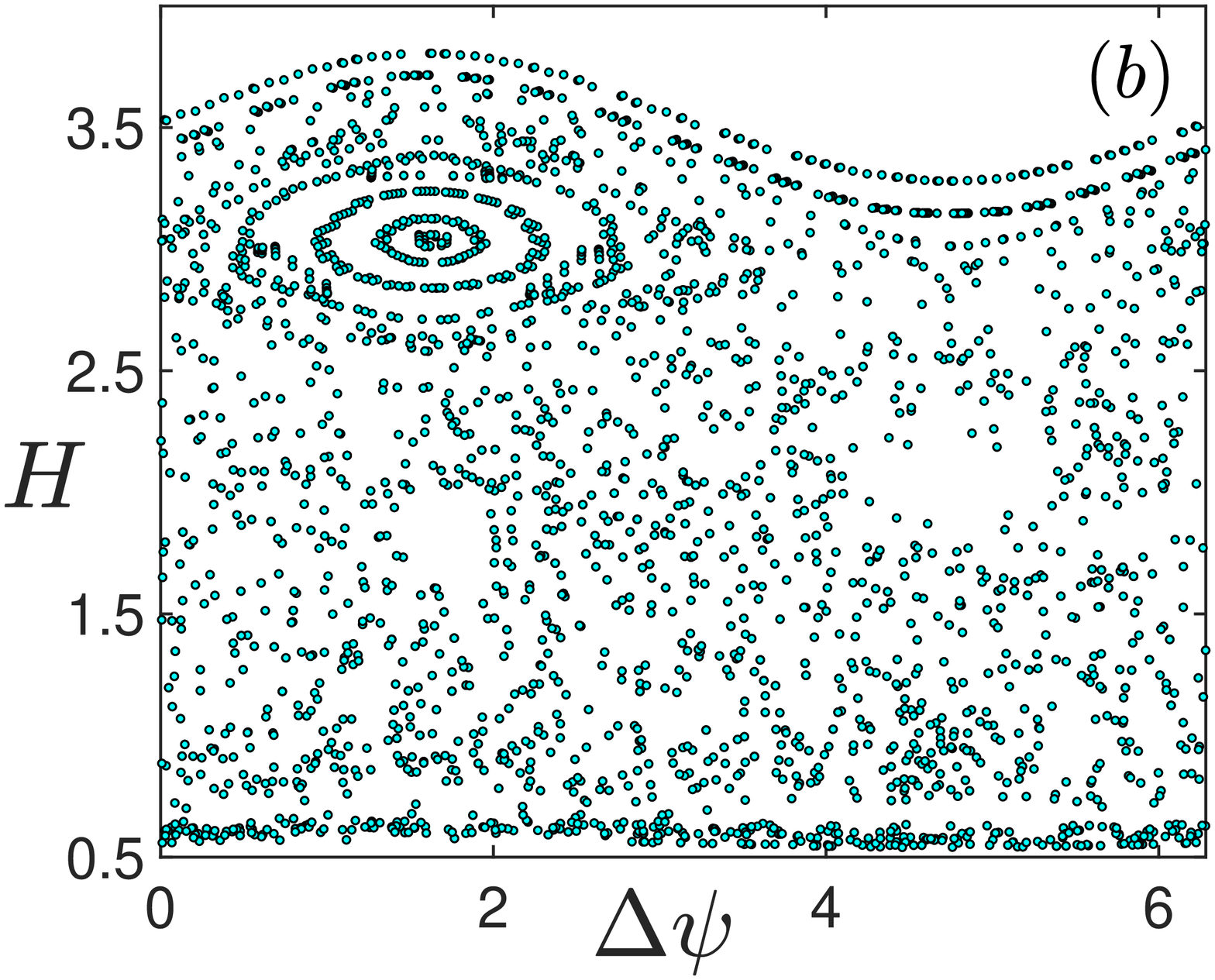}
\end{minipage}}
\caption{Poincar\'e mappings of ($H$, $\psi$) at $\eta=2n\pi+\pi/2$ with the same definition of $\Delta\psi$ with figure~\ref{figChirikov-Ax} for $a=5$, $\mathbf{A}_1=0.2sin(\tau)\mathbf{e}_y~(k_1=1)$, $\bar{P}_x=0$ and (a) $\bar{P}_y=0$; (b) $\bar{P}_y=2$.}
\label{figChirikov-Ay}
\end{figure}

In figure~\ref{figChirikov-Ay} we show the results for the case where the polarization direction of the perturbative laser is orthogonal to that of the dominant one. The parameters are the same with those in figure~\ref{fig-H-a-5-a1x-02-py-0} with different choices of $\bar{P}_y$. It demonstrates that the lower and upper stochastic boundaries are, respectively, in agreement with equation~(\ref{eqH_minimal}) and equation~(\ref{eqH_maxim}) for $\bar{P}_y=0$, and with equation~(\ref{eqH_minimal_py}) and equation~(\ref{eqH_maxim_py}) for $\bar{P}_y\gg a_1$. For all the cases, the maximum electron kinetic energy is consistent with equation~(\ref{eqgamma_max}).

\begin{figure}
\centering
\subfigure{
\label{fig-H-a-5-a1x-02-py-0-vp-11}
\begin{minipage}[bht]{0.3\textwidth}
\includegraphics[width=1\textwidth]{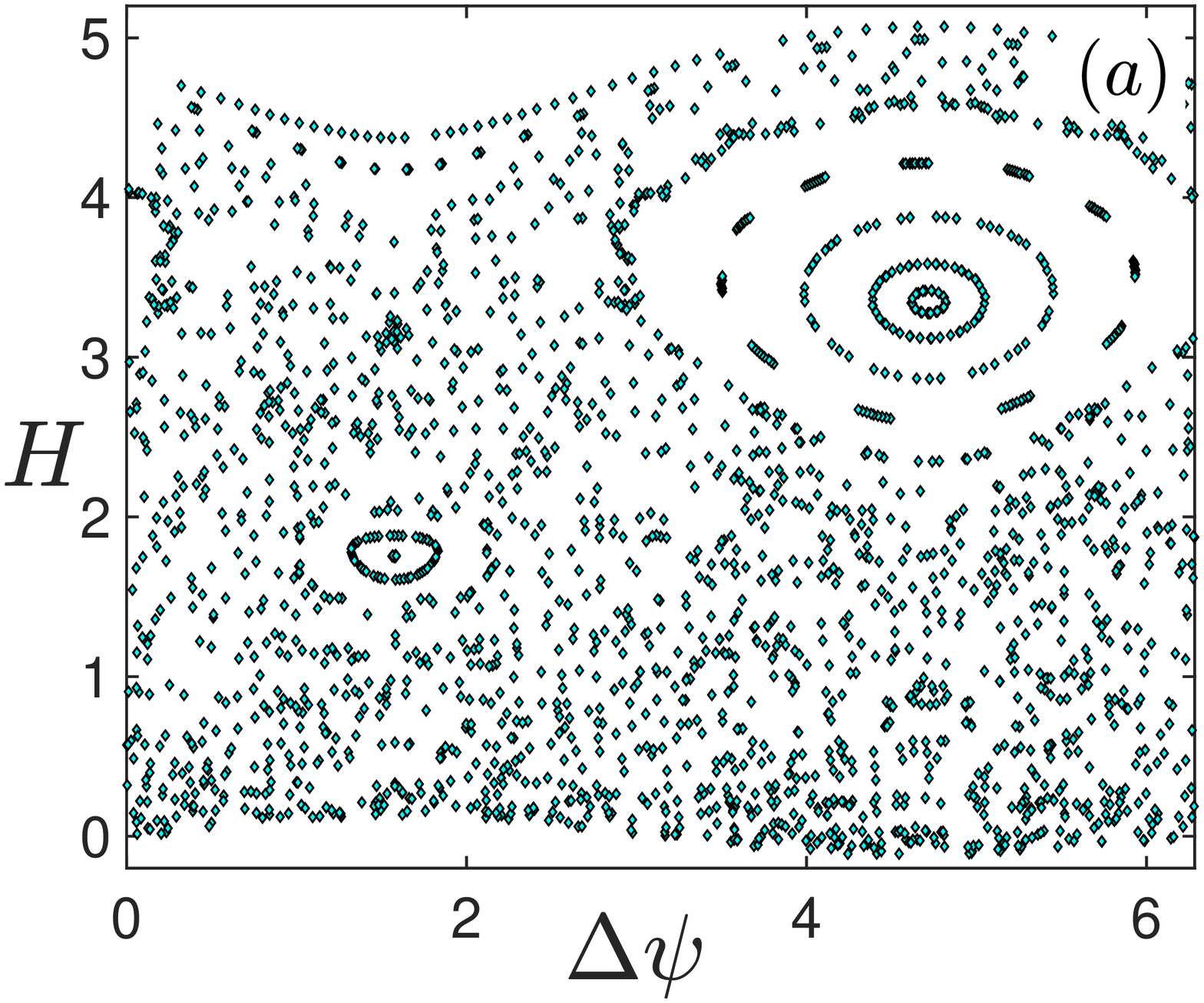}
\end{minipage}}
\subfigure{
\label{fig-H-a-5-a1x-02-py-0-vp-12}
\begin{minipage}[bht]{0.3\textwidth}
\includegraphics[width=1\textwidth]{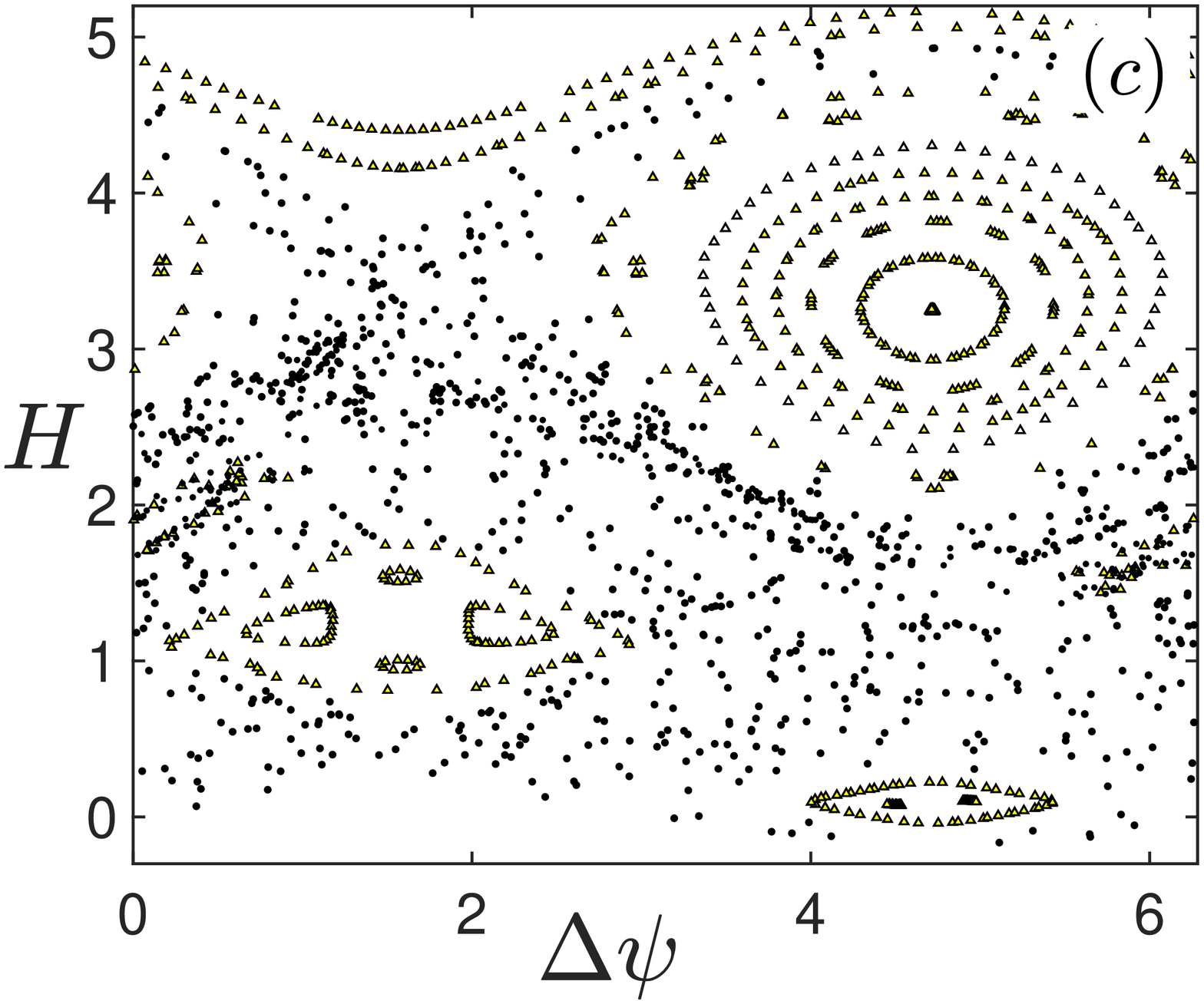}
\end{minipage}}
\subfigure{
\label{fig-H-a-5-a1x-02-py-0-vp-14}
\begin{minipage}[bht]{0.3\textwidth}
\includegraphics[width=1\textwidth]{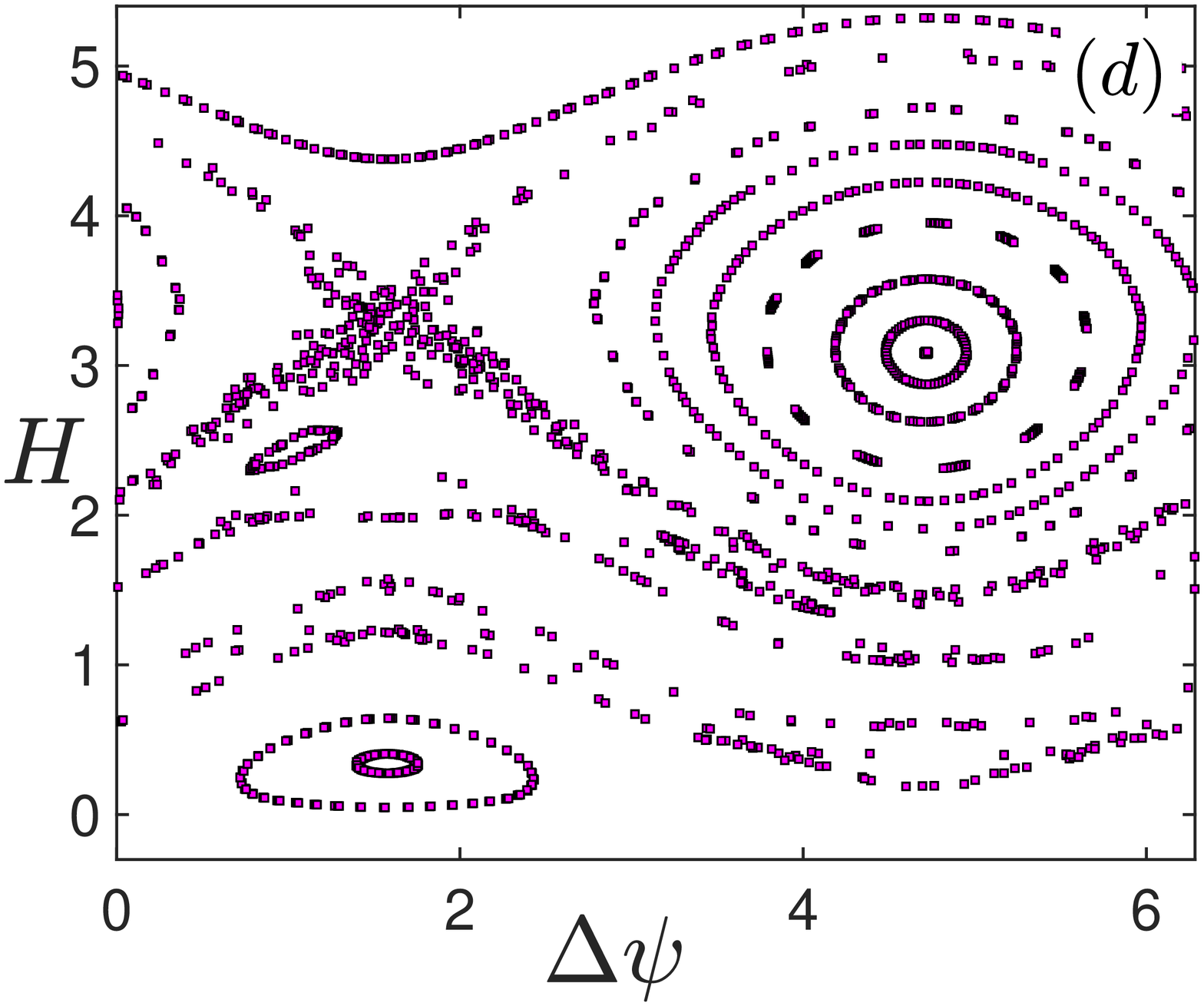}
\end{minipage}}
\subfigure{
\label{fig-gamma-a-5-a1x-02-py-0-vp-11}
\begin{minipage}[bht]{0.3\textwidth}
\includegraphics[width=1\textwidth]{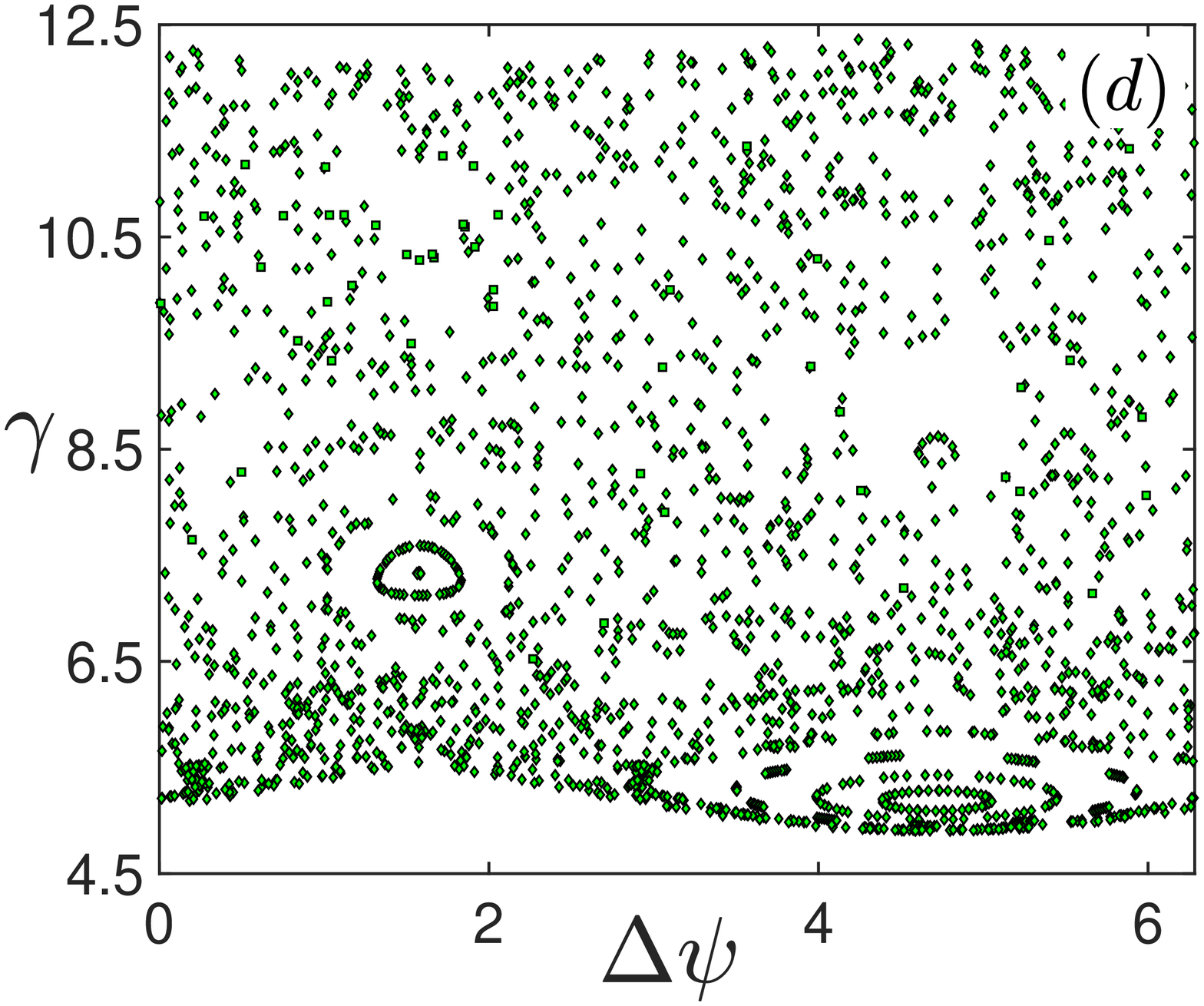}
\end{minipage}}
\subfigure{
\label{fig-gamma-a-5-a1x-02-py-0}
\begin{minipage}[bht]{0.3\textwidth}
\includegraphics[width=1\textwidth]{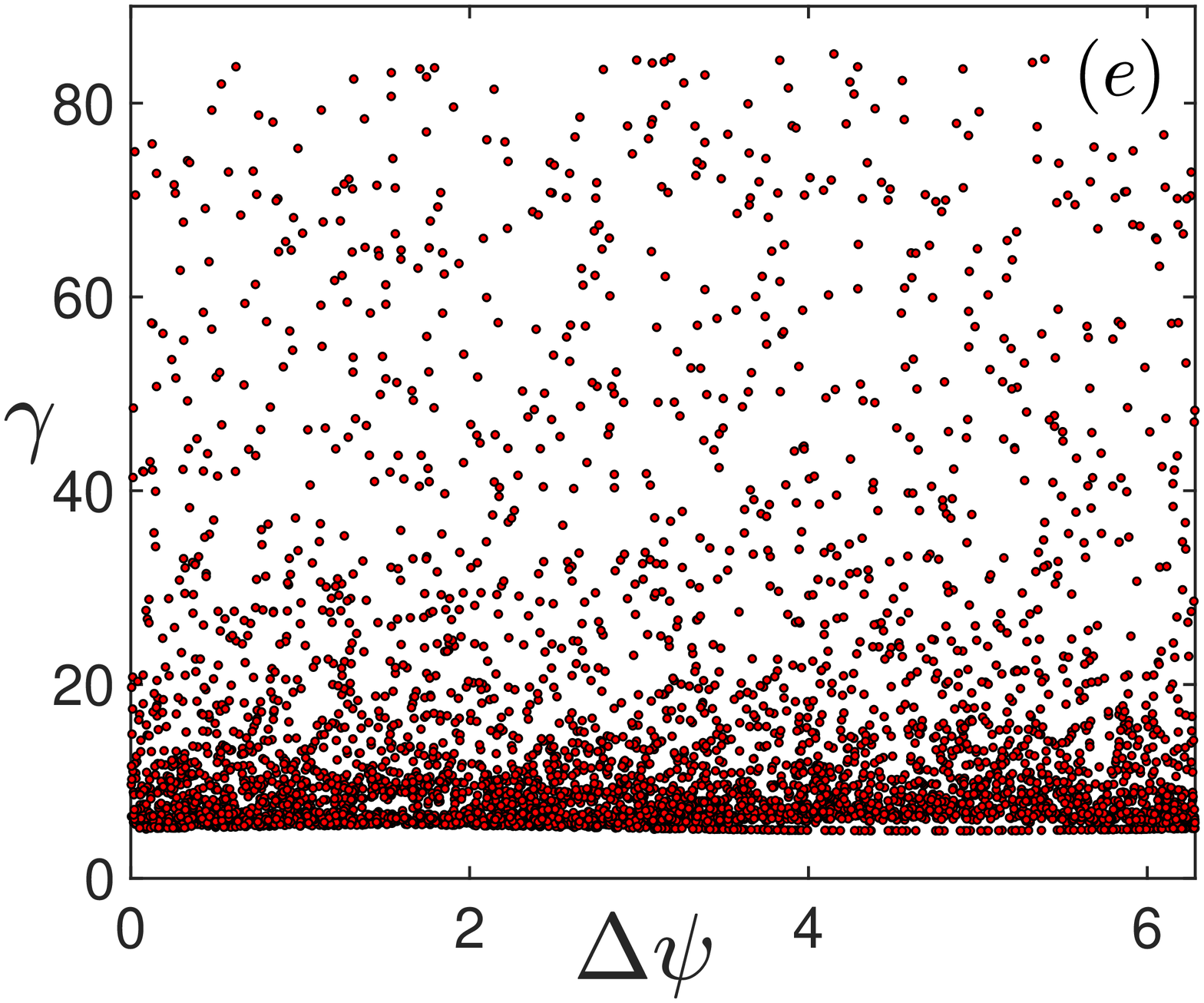}
\end{minipage}}
\subfigure{
\label{fig-H-a-5-a1x-02-py-0-vp-11-tau}
\begin{minipage}[bht]{0.3\textwidth}
\includegraphics[width=1\textwidth]{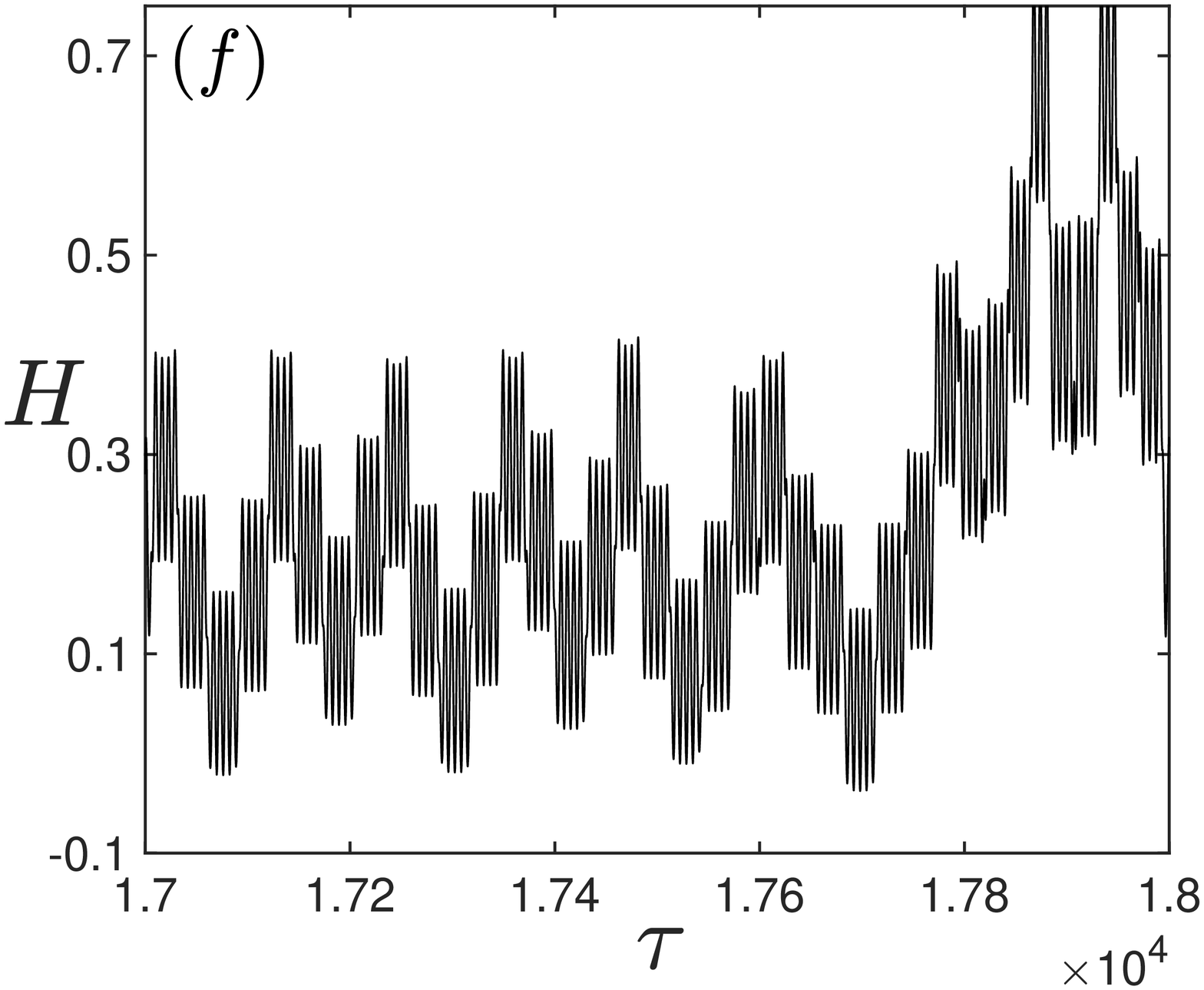}
\end{minipage}}
\caption{The Poincar\'e mappings of ($H$, $\psi$) for the same conditions with figure~\ref{fig-H-a-5-a1x-02-py-0} but for (a) $v_p=1.1$; (b) $v_p=1.2$; and (c) $v_p=1.3$. The mappings of ($\gamma$, $\psi$) are shown in (d) for $v_p=1.1$  corresponding to figure~\ref{fig-H-a-5-a1x-02-py-0-vp-11}; and (e) for $v_p=1$ corresponding to figure~\ref{fig-H-a-5-a1x-02-py-0}. Schematic view of evolution of $H$ corresponding to figure~\ref{fig-H-a-5-a1x-02-py-0-vp-11} is shown in (f).}
\label{figChirikov-Ax-vp}
\end{figure}

Figure~\ref{figChirikov-Ax-vp} shows the results for the same conditions with figure~\ref{fig-H-a-5-a1x-02-py-0} but for superluminal case with different $v_p>1$ as well as the Poincar\'e mapping of ($\gamma$, $\Delta \psi$) corresponding to figure~\ref{fig-H-a-5-a1x-02-py-0} for luminal case. Note that the lower boundary of the stochastic region for luminal case satisfies $H_{min}\ll \sqrt{(v_p^2-1)(1+a^2)}$. Comparing with figure~\ref{fig-H-a-5-a1x-02-py-0}, figure~\ref{fig-H-a-5-a1x-02-py-0-vp-11} shows that, for relatively small $v_p$, the stochastic region in $H$ remains almost unchanged for large $H$ whereas the lower boundary of the stochastic region can extend to negative $H$ with small magnitude. However, some new stability islands appear for the superluminal case. From figure~\ref{fig-H-a-5-a1x-02-py-0-vp-12} (where the stochastic region is denoted by the black dots and the stability islands are depicted by the yellow triangles), we see that the number and scale of these stability islands increase with $v_p$ but the electron in the stochastic region can still reaches $H\approx 0$ for $v_p<v_{ps}$. Further increasing of $v_p$ such that $v_p>v_{ps}$ will eventually terminate the possibility of the stochastic acceleration of electrons ($v_{ps}\approx 1.3$ for these simulations of $a_1=0.2$). Even though the change of the stochastic region in $H$ for $v_p<v_{ps}$ is small, the maximum electron kinetic energy is significantly decreased by the superluminal phase velocity (e.g., see figure~\ref{fig-gamma-a-5-a1x-02-py-0-vp-11} and \ref{fig-gamma-a-5-a1x-02-py-0} for $v_p=1.1$ and $v_p=1$, respectively). Recalling that $H\approx 0$ is accessible as long as stochastic region exists and $\gamma_{max}$ is insensitive to small $H$, the maximum electron kinetic energy (e.g., see figure~\ref{fig-gamma-a-5-a1x-02-py-0-vp-11}) can be well predicted by equation~(\ref{eqgamma_max_super}). In figure~\ref{fig-H-a-5-a1x-02-py-0-vp-11-tau} we sketched the evolution of $H$ for $v_p=1.1$, which confirms that for small $H\ll \sqrt{(v_p^2-1)(1+a^2)}$, the electron oscillation period remains almost unchanged unlike the luminal case shown in figure~\ref{fig-H_diff-chi}. We can also see the large variation of $H$ due to the kicks and approximately constant large oscillation of $H$ in the adiabatic region, which is consistent with the estimates in last section.

\section{Conclusions}
\label{sec_conclusions}
In conclusion, we consider electron dynamics in the fields of colliding laser beams. It shows that the proper choice of canonical variables and effective time, such that the new Hamiltonian is conserved for electrons in a dominant laser field, greatly simplifies analytical treatment of the problem and allows us to clearly reveal the physical picture of stochastic electron dynamics and find the maximum electron kinetic energy. For the case of counter propagating planar laser beams and dominant laser with relativistic intensity, $a > 1$, we find that when the amplitude of perturbative laser ($a_1 < a$) exceeds the threshold value, stochastic acceleration of electrons becomes possible within some range of $H$ and thus electron kinetic energy. The maximum electron kinetic energy, which could be gained under stochastic acceleration, can significantly exceeds the ponderomotive scaling for electron in the dominant laser only. 

For the case of colliding laser waves polarized in the same direction, the presence of initial condition $\bar{P}_y\neq 0$ will increase the threshold value as shown in equation~(\ref{eqThreshold-a2}) and narrow the stochastic region, provided that it only enlarges the effective electron mass and thus the impact of the perturbative laser becomes weaker. On the other hand, the presence of $|\bar{P}_x|\tilde{<}a$ could decrease the threshold value as shown in figure~\ref{fig-f-m-px} but $|\bar{P}_x|\gg a$ has the opposite effect. This is a competing result of increasing the effective electron mass and electron momentum along the dominant laser electric field. However, $\bar{P}_x$ will slightly increase the lower boundary, $H_{min}^x$, of the stochastic region and thus will decrease the ratio of the maximum kinetic energy $\gamma_{max}\approx E_p/2H_{min}^x$ over the ponderomotive scaling $E_p/H_0$. 

For the case where the polarization direction of the perturbative laser is orthogonal to that of the dominant laser, the threshold value in equation~(\ref{eqThreshold-a2-Ay}) is much larger than that for lasers being parallel polarized in equation~(\ref{eqThreshold-a2}) for $\bar{P}_y=0$. However, the presence of $\bar{P}_y$ could decrease this threshold as shown in equation~(\ref{eqThreshold-a2-Ay-Py}) when $a_1<|\bar{P}_y|\tilde{<} a^{3/8}$ and even to the value comparable to that in equation~(\ref{eqThreshold-a2}) when $|\bar{P}_y|>1$. We find that, regardless the orientation of the perturbative laser, its amplitude $a_1$, as long as it is above the threshold value for stochasticity, has a weak impact on the lower boundary of the stochastic region and thus the maximum electron kinetic energy for $\bar{P}_x=0$.  

The impact of superluminal phase velocity $v_p>1$ on stochastic electron dynamics is qualitatively discussed in section \ref{sec_superl}. It shows that the stochastic region for $H> \sqrt{(v_p^2-1)(1+a^2)}$ is not affected by the superluminal phase velocity, whereas for $H\ll \sqrt{(v_p^2-1)(1+a^2)}$, both the variation of $H$ due to the kick and the derivative of the time interval between two consecutive kicks with respect to $H$ are approximately constant such that there exists a threshold value of $v_p$, i.e., $v_{ps}$. When $v_p<v_{ps}$, the lower boundary is extended to small negative $H$ and new stability islands appear with increasing $v_p$. Although the change of the stochastic region in $H$ is small, the maximum electron kinetic energy is significantly decreased by the superluminal phase velocity. On the other hand, when $v_p>v_{ps}$, the stochasticity is terminated and thus the corresponding electron kinetic energy is rather limited. 

The Hamiltonian equations are numerically integrated, whose results shown in figure~\ref{figChirikov-Ax}-\ref{figChirikov-Ax-vp} are in a very good agreement with the findings from our analytic theory. We notice that the approach presented in this letter could be applied to many other cases including electron dynamics in the laser and quasi-stationary electromagnetic fields, in intense laser and Langmuir waves, etc.

This work was supported by the University of California Office of the President Lab Fee grant number LFR-17-449059.

\appendix
\section{}
\label{AppendixVmn}

The stochastic condition for the case of $\mathbf{A}_1=a_1sin(k_1\tau)\mathbf{e}_x$ and general $\bar{P}_x$ can be found from the point of view of resonant islands overlapping, $\bar{K}=(\delta \omega+\delta \omega')/2\Delta \omega>1$. For this purpose,
the unperturbed electron motion can be expressed by using the action-angle variables ($I$ and $\theta$):
\begin{equation}
I=-\frac{2\bar{P}^2+a^2}{2H_0}~,\textup{and}~\theta=\hat{\eta}-\frac{a^2sin(2\hat{\eta})}{2(2\bar{P}^2+a^2)}-\frac{4a\bar{P}_x\left[cos(\hat{\eta})-1\right]}{2\bar{P}^2+a^2},\label{eqActionandTheta}
\end{equation}
where $\hat{\eta}=\eta-2n\pi$ and the electron oscillating frequency in equation~(\ref{eqOmega}) can be written in terms of $I$ as $\omega(I)=(2\bar{P}^2+a^2)/2I^2$. Given that the electron motion is periodic with $\theta$, we can expand the first order correction to $H$ in equation~(\ref{eqHamiltonian_luminal}) in Fourier series: \begin{equation}
H_1=2a_1\left[asin(\eta)+\bar{P}_x\right]sin(k_1\xi)/\chi=\sum_{m,n}V_{mn}(I)e^{i(m\theta-nk_1\xi)}+c.c.,\label{eqFourier}
\end{equation}
and $n=\pm 1$ as seen from equation~(\ref{eqHamiltonian_luminal}). As a result, the resonance, corresponding to a constant phase of the perturbation, occurs at $\omega(I)=nk_1/m$. 

The Fourier coefficients $V_{mn}$ in equation~(\ref{eqFourier}) are given by 
\begin{equation}
V_{mn}=\frac{k_2}{(2\pi)^2}\int_0^{2\pi/k_2}\int_0^{2\pi}H_1(I,\theta,\xi)e^{-i(m\theta-nk_2\xi)}d\theta d\xi.\label{eqAdd_Vmn}
\end{equation}
After some algebra, we arrive at 
\begin{equation}
\left|V_{mn}\right|=\frac{a}{2I}e^{-imC}\sum_{h=0,\pm 1}\left[ha+2\delta_h^0\bar{P}_x\right]C_{m-h}\left [\frac{ma^2}{2(2\bar{P}^2+a^2)},\frac{4ma\bar{P}_x}{2\bar{P}^2+a^2}\right],\label{eqAdd_Vmn_general}
\end{equation}
where 
\begin{equation}
 C_N(\alpha,\beta)=\sum_{q=-\infty}^{\infty}J_{q}(\alpha)J_{N-2q}(\beta)i^{N-2q},\label{eqAdd_general_bessel}
\end{equation}
is the generalized Bessel function \cite{rax1992compton,nikishov1964ai}, $C=4a\bar{P}_x/(2\bar{P}^2+a^2)$, and $\delta_i^j$ is the Kronecker symbols. Notice that similar result for $V_{mn}$ was obtained in \cite{rax1992compton} by using multidimensional Hamiltonian methods.

The width of the island is approximated \cite{rechester1979stochastic} as
\begin{equation}
\delta \omega=4\left|2V_{mn}\frac{d\omega}{dI}\right|^{1/2},\label{eqDelta_omega}
\end{equation}
whereas the spacing between possibly overlapping resonances is
\begin{equation}
\Delta \omega=|\omega(I_{m'})-\omega(I_m)|\approx \omega^2/k_1,\label{eqdelta_omega}
\end{equation}
for $|m|\gg 1$. Then the stochastic condition reads 
\begin{equation}
\bar{K}^2=a_1\frac{16m^2}{(2\bar{P}^2+a^2)}\sum_{h=0,\pm 1}\left[ha+2\delta_h^0\bar{P}_x\right]C_{m-h}\left [\frac{ma^2}{2(2\bar{P}^2+a^2)},\frac{4ma\bar{P}_x}{2\bar{P}^2+a^2}\right]>1.\label{eqSquare_of_K-Append}
\end{equation}

\bibliography{main.bib}% Produces the bibliography via BibTeX.

\end{document}